\documentclass[a4paper,prd,unsortedaddress,superscriptaddress,showpacs,twocolumn,floatfix]{revtex4}
\usepackage{pstricks}
\usepackage{graphicx}
\usepackage{bm}
\newcommand{\xipp}{\ensuremath{\xi''}}
\newcommand{\lln}{Institut de Recherche en Math\'ematique et Physique, 
  Universit\'e catholique de Louvain, B--1348 Louvain--la--Neuve, Belgium} 
\newcommand{\ethz}{Institute for Particle Physics, Eidgen\"ossische Technische
Hochschule Z\"urich, CH--8093 Z\"urich, Switzerland}
\newcommand{\PSI}{Paul Scherrer Institute, CH--5232 Villigen--PSI, Switzerland}
\newcommand{\lpcc}{LPC-Caen, ENSICAEN, Universit\'e de Caen Basse-Normandie,
CNRS/IN2P3-ENSI, F--14050 Caen, France}
\newcommand{\msu}{National Superconducting Cyclotron Laboratory and Department
of Physics and Astronomy, Michigan State University, East Lansing, 48824 MI, USA}
\newlength{\textlarg}

\begin{document}
%%%%%
\title{Measurement of the Michel Parameter \xipp{} in Polarized Muon Decay
and Implications on Exotic Couplings of the Leptonic Weak Interaction}

%%%%%
%\affiliation{\lln}
%\affiliation{\ethz}
%\affiliation{\PSI}
%\affiliation{\lpcc}
%%%%%

%%% quote always \altaffiliaton after \affiliation

\author{R.~Prieels}
\affiliation{\lln}

\author{O.~Naviliat-Cuncic}
%\altaffiliation[Present address: ]{NSCL and Department of Physics and Astronomy,
%Michigan State University, East Lansing, 48824 MI, USA}
\altaffiliation[Corresponding author: ]{naviliat@nscl.msu.edu}
\affiliation{\ethz}
\affiliation{\lpcc}
\affiliation{\msu}

\author{P.~Knowles}
\altaffiliation[Present address: ]{Rilkeplatz 8/9, 1040 Vienna, Austria}
\affiliation{\lln}

\author{P.~Van~Hove}
\altaffiliation[Permanent address: ]{Institut Pluridisciplinaire
Hubert Curien, 67037--Strasbourg, France}
\affiliation{\lln}

\author{X.~Morelle}
\altaffiliation[Present address: ]{Goodyear and Dunlop tires, Luxemburg}
\affiliation{\ethz}

\author{J.~Egger}
\affiliation{\PSI}

\author{J.~Deutsch}
\altaffiliation[Deceased.]{}
\affiliation{\lln}

\author{J.~Govaerts}
\affiliation{\lln}

\author{W.~Fetscher}
\affiliation{\ethz}

\author{K.~Kirch}
\affiliation{\ethz}
\affiliation{\PSI}

\author{J.~Lang}
\affiliation{\ethz}

\date{\today}
%\date{..., 2006}

%%%%%%%%%%%%%%%%%%%%%%%%%%%%%%%%%%%%%%%%%%%%%%%%%%%%%%%%%%%%%%%%%%%%%%%%%%%%
\begin{abstract}
The Michel parameter \xipp{} has been determined from a measurement of
the longitudinal polarization of positrons emitted in the decay of
polarized and depolarized muons.
The result, $\xipp = 0.981\pm 0.045_{\rm stat}\pm 0.003_{\rm syst}$,
is consistent with the Standard
Model prediction of unity, and provides an order of magnitude improvement
in the relative precision of this parameter.
This value sets new constraints on exotic couplings beyond the
dominant $V-A$ description of the leptonic weak interaction.
\end{abstract}

\pacs{12.60.Cn, 13.35.Bv, 13.88.+e, 14.60.Ef}% PACS codes

% Use showpacs class option to display
\maketitle
%%%%%%%%%%%%%%%%%%%%%%%%%%%%%%%%%%%%%%%%%%%%%%%%%%%%%%%%%%%%%%%%%%%%%%%%%%%%
\section{Introduction}

Normal muon decay, $\mu^+ \rightarrow e^+ \nu_e \bar{\nu}_{\mu}$, is
a sensitive elementary process to probe the Lorentz structure of the
charged current sector and to search for new physics beyond
the Standard Model (SM) of electroweak interactions~\cite{Herczeg86,Kuno01}.
Assuming a local, derivative free, four-fermion point interaction,
invariant under Lorentz transformations, the effective muon decay
amplitude can be expressed at lowest order as~\cite{Fetscher86}
\begin{equation}
{\cal M} = \frac{4 G_F}{\sqrt{2}} \sum_{\gamma=S,V,T \atop \epsilon,\mu=R,L} 
g^{\gamma}_{\epsilon\mu}
\langle \bar{e}_{\epsilon} | \Gamma^{\gamma} | (\nu_e)_n \rangle
\langle (\bar{\nu}_\mu)_m | \Gamma_{\gamma} | \mu_{\mu} \rangle
\label{eq:mumatrix}
\end{equation}
where $G_F$ is the Fermi constant and $g^{\gamma}_{\epsilon\mu}$ are
the couplings associated with the scalar, vector, and tensor
interactions characterized by the operators $\Gamma^\gamma$.
Each interaction term involves electrons of chirality $\epsilon$ and
muons of chirality $\mu$, whereas the indices $n$ and $m$ indicate the
chiralities of the neutrinos which are uniquely determined once
$\gamma$, $\epsilon$ and $\mu$ are fixed.
Neutrino masses are here neglected.
Within the SM, $g^V_{LL} = 1$, and all other couplings are zero.

Observables in muon decay are conveniently expressed in terms of the Michel
parameters~\cite{Michel50} which are bilinear combinations of the couplings
$g^{\gamma}_{\epsilon\mu}$~\cite{Fetscher95}.
Most Michel parameters are known with uncertainties below the percent
level~\cite{PDG12}.
In particular, new results have recently been reported on the
parameters ${\cal P}^\pi_\mu \xi$~\cite{Bueno11} and
$\rho$, $\delta$~\cite{Hillairet12},
for which the total errors reach respectively $(+16.8,-6.9)\times10^{-4}$,
$2.6\times 10^{-4}$, and $3.4\times 10^{-4}$.
A notable exception among Michel parameters is \xipp, which
characterizes the angular and energy dependence of the positron
longitudinal polarization in polarized muon decay.
This parameter has been determined only once~\cite{Burkard85a,Burkard85b},
yielding $\xipp = 0.65\pm 0.36$, where the error is dominated by
statistics.

We report here the results of an improved determination of the Michel
parameter \xipp{} deduced from a measurement of the longitudinal
polarization of positrons emitted from decays of both highly polarized
and depolarized muons, and discuss the implications of such a measurement
in constraining exotic couplings beyond the SM that could contribute to the muon
decay amplitude.

%%%%%%%%%%%%%%%%%%%%%%%%%%%%%%%%%%%%%%%%%%%%%%%%%%%%%%%%%%%%%%%%%%%%%%%%%%%%
\section{The Longitudinal Polarization}

Using the standard formalism to express the muon decay
rate~\cite{Fetscher95,PDG12}, assuming the SM values $\delta = \rho = 3/4$,
neglecting the mass of the positron and the contribution of
radiative corrections~\cite{Mehr79}, the longitudinal
polarization $P_L$ of positrons emitted with reduced energy $x$ at an
angle $\theta$ relative to the 
direction of the oriented muon spins (with polarization $P_\mu$) can be expressed as
\cite{Scheck78}
\begin{equation}
P_L(P_\mu,x,z) = {\xi'}\left[
1 + k(P_\mu,x,z)~\Delta \right]
\label{eq:longpol}
\end{equation}
where $z=\cos{\theta}$, $\xi'$ is the SM expectation for the positron longitudinal
polarization, $k(P_\mu,x,z)$ is an enhancement factor and $\Delta$ is
the combination of Michel parameters
\begin{equation}
\Delta \equiv (\xipp/\xi\xi' - 1) \approx (\xipp - 1)\,.
\label{eq:delta}
\end{equation}

The enhancement factor $k(P_\mu,x,z)$ determines the sensitivity to
\xipp{} embedded in $\Delta$ and is given by
\begin{equation}
k(P_\mu,x,z) = 
\frac{P_\mu z \xi (2x-1)}{(3-2x)+P_\mu z \xi (2x-1)}\,.
\label{eq:enhfact}
\end{equation}
The reduced total energy of the positron, $x=E_e/W_e$, is
normalized to the decay endpoint $W_e=52.83$~MeV\@.

\begin{figure}[ht]
\centerline{\includegraphics[width=1.2\linewidth]{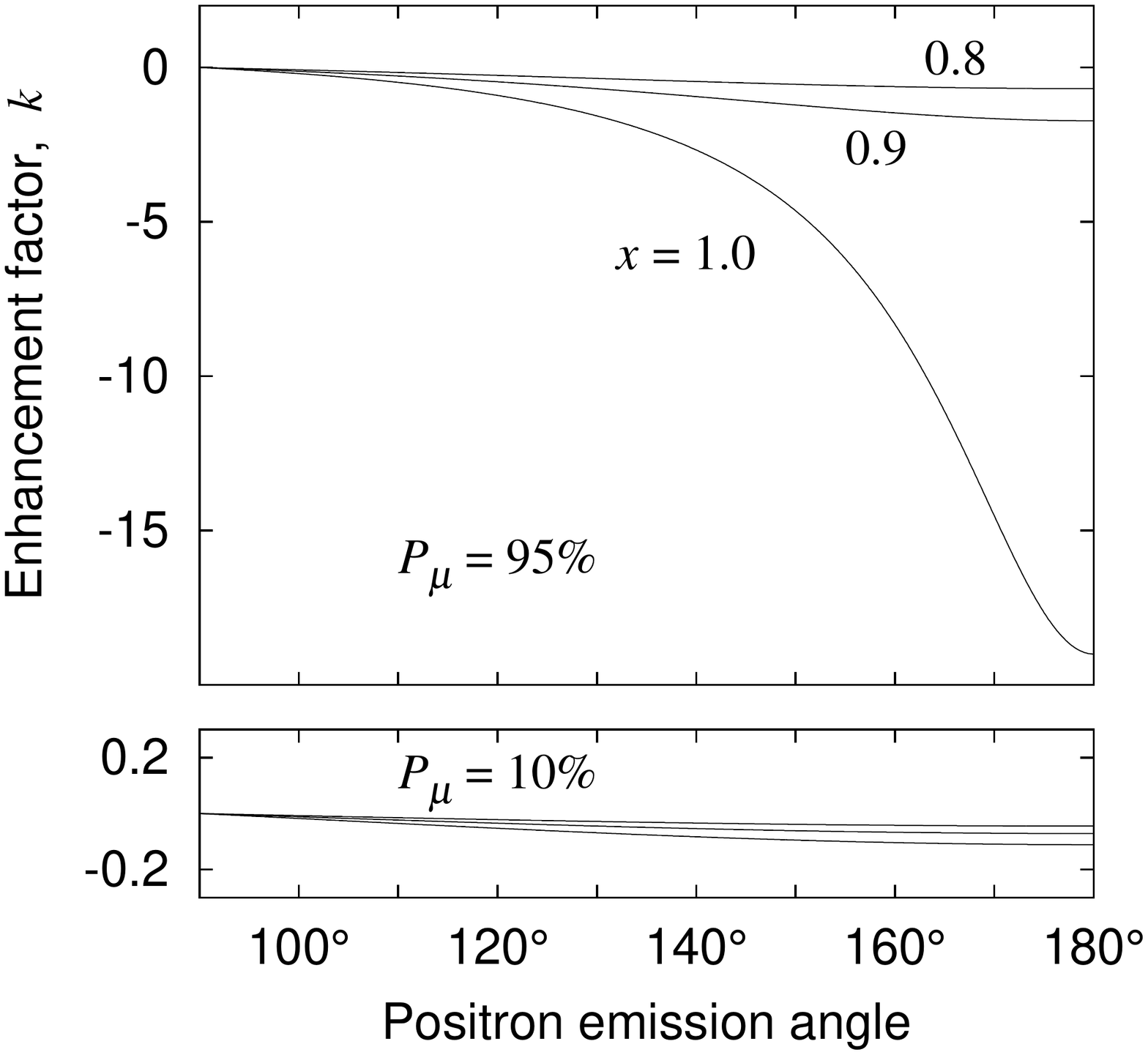}}
\vspace*{-2 mm}
\caption{Variation of the enhancement factor $k(P_\mu,x,\cos{\theta})$
  as a function of the positron emission angle, $\theta$, for three
  values of the reduced energy $x$. Upper panel: for $P_\mu =
  0.95$. Lower panel: for $P_\mu = 0.10$.}
\label{fig:enhfac}
\end{figure}

In the SM, the Michel parameters assume values of $\xi = \xi' =
\xipp = 1$ so that $\Delta = 0$ and the electron longitudinal
polarization has no energy nor angular dependence.
For positrons emitted from highly polarized muons, with energies close
to the end-point and at backward angles relative to the muon spin, the
enhancement factor takes on large values.
For illustration, Fig.~\ref{fig:enhfac} shows the behavior of the enhancement factor
for two values of the muon polarization, $P_{\mu} = 0.95$ and $P_{\mu}
= 0.10$, assuming $\xi = 1$.
The dependence as a function of the variables $P_{\mu}$, $x$, and
emission angle $\theta$ indicates the most favorable kinematic conditions
in order to achieve a large experimental sensitivity to \xipp.

%%%%%%%%%%%%%%%%%%%%%%%%%%%%%%%%%%%%%%%%%%%%%%%%%%%%%%%%%%%%%%%%%%%%%%%%%%%%
\section{Experimental setup}

\begin{figure}[t]
\centerline{\includegraphics[width=\linewidth]{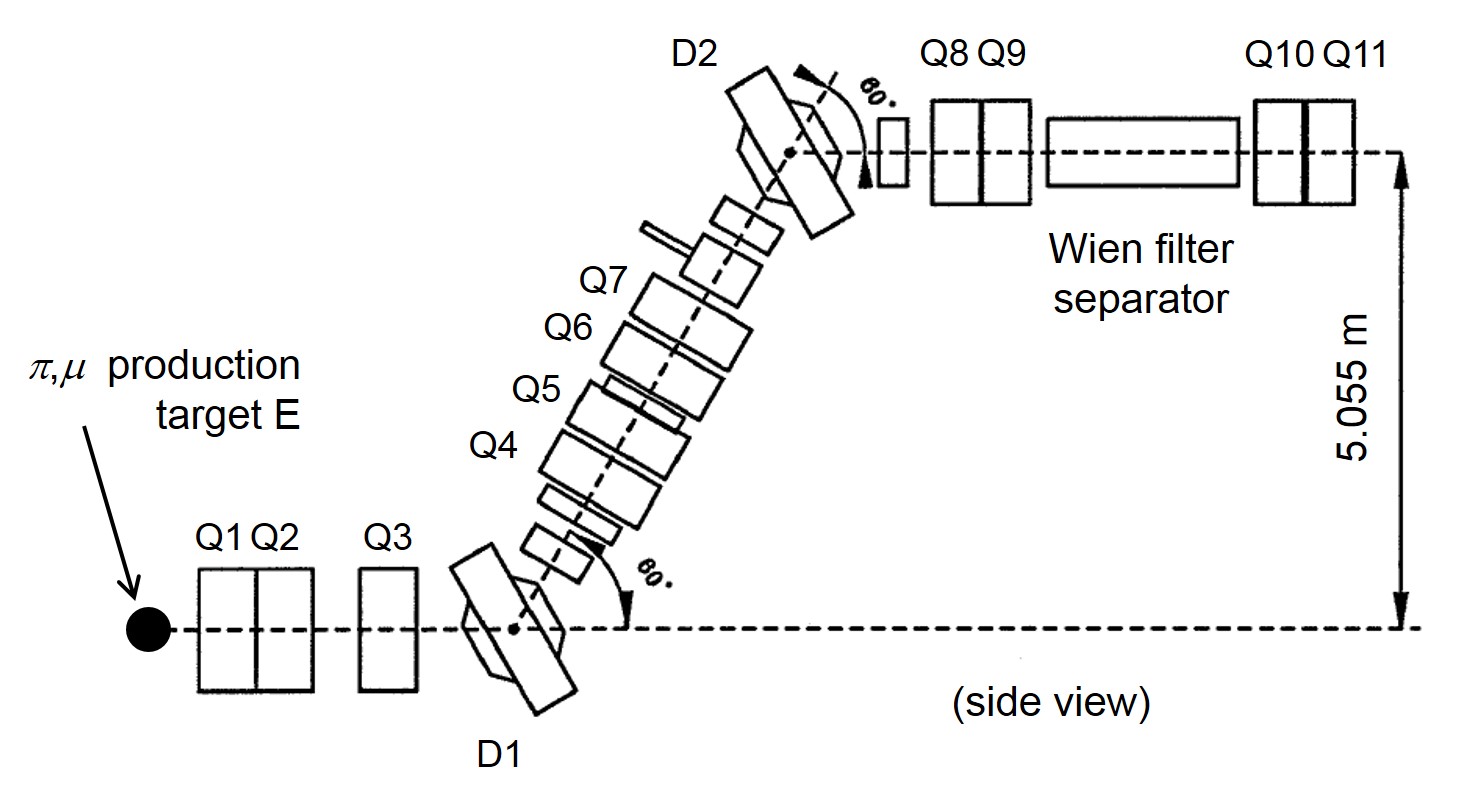}}
\vspace*{-2 mm}
\caption{Vertical profile of the $\pi$E3 beam-line at the
  Paul Scherrer Institute showing the quadrupoles, the dipoles and the Wien
filter.} 
\label{fig:piE3}
\end{figure}

The measurement has been carried out at the $\pi$E3 beam-line of the
Paul Scherrer Institute, Villigen, Switzerland.
A layout of the magnetic elements of the beam-line is shown in
Fig.~\ref{fig:piE3}. The elements are located on a vertical bending plane
so that the extracted beam is 5~m higher than the primary proton beam
and the muon production target.
The beam line includes two bending dipoles and a series of quadrupoles. An important
element for the beam purification is the velocity (Wien) filter for the
separation between muons and positrons in the beam.

%%% THIS IS EXPLANED IN THE SUMMARY, no need to state it here
%%%
%The data set
%presented in this work was collected during a six week long run which took
%place in 2001.
%

%%%%%%%%%%%%%%%%%%%%%%%%%%%%%%%%%%%%%%%%%%%%%%%%%%%%%%%%%%%%%%%%%%%%%%%%%%%%
\subsection{Muon beam}

The elements of the beam-line were tuned to select particles with
momentum of 28.5~MeV$/c$, thus favoring the surface muons created
from pions decaying at rest in the carbon production target.
These muons are naturally backward polarized relatively to their
emission direction.
The initial positron contamination of the beam at this momentum was
eight times larger than the muon beam intensity.
The~Wien filter (Fig.~\ref{fig:piE3})
was used to separate $e^+$ from $\mu^+$, diverting positrons
from the main beam axis.
At the muon implantation target (see below) the centroid of the $e^+$
beam profile was measured to be 16~cm away from the beam axis, far
enough
from the acceptance aperture of the spectrometer.
After separation, the resulting $e/\mu$ contamination at the
implantation target was finally 12.5\%.
The intensity of the muon beam was $3\times10^7~\mathrm{s}^{-1}$ for a
typical primary proton beam of 1.6~mA\@.
After the end of the beam-line, muons were transported in air up to
the implantation target over a distance of 35~cm.

%%%%%%%%%%%%%%%%%%%%%%%%%%%%%%%%%%%%%%%%%%%%%%%%%%%%%%%%%%%%%%%%%%%%%%%%%%%%
\subsection{Muon polarization and implantation targets}
\label{sec:Pmu}

A detailed study of the residual muon polarization for muons extracted
from the $\pi$E3 beam-line and
implanted in a thin aluminum (Al) target was carried out in a dedicated
experiment, using the technique based on the Hanle
signal~\cite{Possoz79}. The signal was deduced from the rates of decay
positrons measured by three plastic scintillator
telescopes, as a function of the beam momentum between 25 and
40~MeV$/c$.
At 34~MeV$/c$, cloud muons from pion decays in flight were found to
have a polarization of $-0.24$ relative to surface
muons~\cite{VanHove00}.
The extrapolation of the measured cloud muon intensity and polarization towards
lower momenta results in an effective beam polarization of $P^{\mathrm{Al}}_\mu
= 0.944(11)$ at 28.5~MeV$/c$ for muons implanted in the
polarization-maintaining Al target~\cite{VanHove00}.
The polarization was found to be slightly larger at this momentum than
at the nominal 29.8~MeV$/c$ for surface muons, probably due to energy
losses in the production target.
For this experiment, it is important to work at the largest possible
polarization so the beam momentum was chosen to be 28.5~MeV$/c$.

In another preparatory experiment, various materials were tested for their
ability to depolarize the stopped muons.
Sulfur (S) was chosen since it showed strong depolarization 
%close to 90\% 
%
% PEK comment:
% without defining how `depolarization' is measured, we shouldn't try
% to give a quite confusing number here.  We give P_\mu in the S target
% later in the paper.  Sufficient here to say that S destroys polarization.
%
for muons implanted with a momentum of 28.5~MeV$/c$.

% As this  could be target quality dependent, we decided to extract the
% effective muon residual polarization $P_\mu(S)$  from the presently
% data obtained in the final target setup.
%%%
%0.05(2)$ \cite{VanHove00}.
%%%
Following the results of these tests, two implantation targets
($10\times10~\mathrm{cm}^2$) of the same mass thickness
($0.405~\mathrm{g}/\mathrm{cm}^2$) were used in the final experiment,
with the Al target preserving the muon polarization and the S target
destroying the polarization.
The targets were located in air,
in a residual longitudinal stray field of about 0.1~T, generated by the
magnets of the spectrometer.
That field maintains the muon spin along the positron spectrometer
axis.
%for muons in the polarization maintaining target.

% FIXME: Jan G's comment
% I believe some reference at least, or some further explanation is
% deserved as to how that value was determined.
The transport of muons inside the Wien filter separator affects the
muon spins by a rotation of about $7^\circ$ relative to the beam axis
\cite {Foroughi2000}
%{\bf as calculated according to 
%\ldots (Does anyone remember where
  %that number came from?  Was this a simple model or a full BMT
 % (Phys.Rev.Lett. 2, 435 (1959)) calculation?)},
reducing thereby the average polarization along the 
spectrometer axis to $P_\mu^{\mathrm{Al}} = 0.937(11)$.

Three plastic scintillator telescopes, two located at $\pm90^\circ$
and one at $121^\circ$ relative to the beam direction (none is shown in
Fig.~\ref{fig:spectro}), continuously monitored the positron rate from
the muon implantation targets.
The targets were tilted around their vertical axis with their planes
making an angle of 
$75^\circ$ relative to the beam direction to reduce the shadowing
towards the telescopes.

The dedicated muon polarization experiments clarified both the
choice of the target materials and the selection of the optimal beam
momentum. However, during the main experiment, the actual residual
muon polarization in both targets was
determined from the shape of the measured energy spectrum, as 
discussed in Sec. \ref{sec:PS}.
%
%: from the positron rates as
%measured by the plastic telescopes; from the positron rates
%transmitted through the spectrometer and normalized to telescopes;
%and from the shape of the energy distribution of the transmitted
%positrons.

%%%%%%%%%%%%%%%%%%%%%%%%%%%%%%%%%%%%%%%%%%%%%%%%%%%%%%%%%%%%%%%%%%%%%%%%%%%%
\subsection{Positron spectrometer}

\begin{figure}[ht]
\centerline{\includegraphics[width=\linewidth]{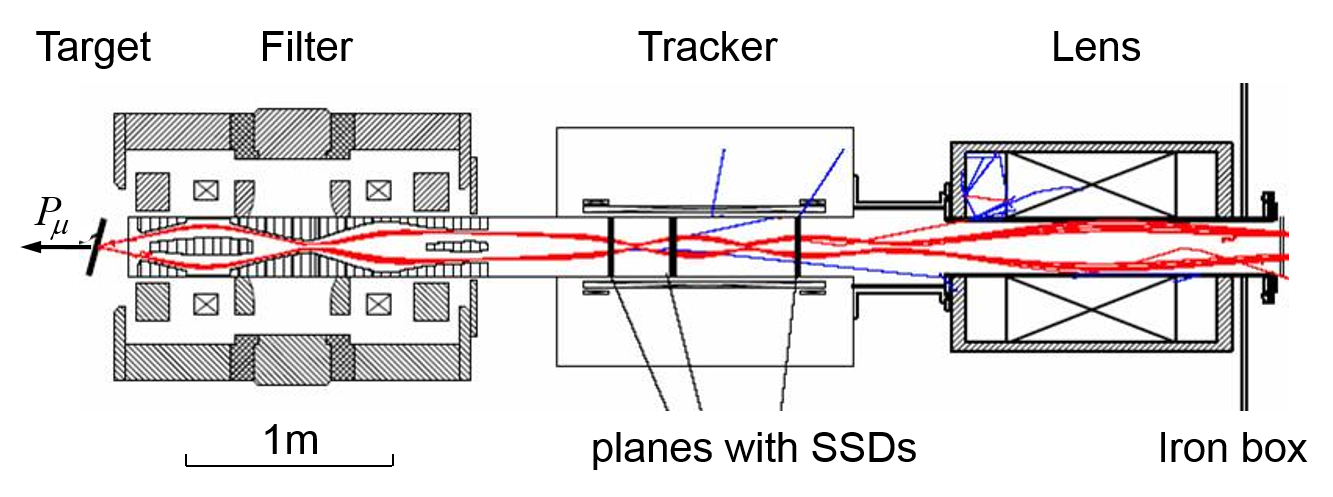}}
\caption{(Color online) General view of the three magnets that are
  assembled into the spectrometer.  The Filter accepts only high
  energy decay positrons, stopping the remainder.  The Tracker is used
  for energy analysis, and the Lens transports the positrons to the
  polarimeter (Fig.~\ref{fig:polar}).
  The red lines are tracks for 50~MeV$/c$ positrons generated via
  Monte-Carlo simulation and the blue lines are secondary particles.
  The arrow on the implantation target shows the average direction of
  the muon polarization.}
\label{fig:spectro}
\end{figure}

The first section of the apparatus is the positron spectrometer
(Fig.~\ref{fig:spectro})
which is located between the implantation target and the polarimeter.
It is composed of three main parts:
a first magnet (Filter) selecting positrons near the
end-point energy, a second magnet (Tracker) where the positron
momentum is measured, and a third magnet (Lens) which focuses the
positrons into the polarimeter.
%
%% All three magnets were recycled from other experiments.

Extensive Monte-Carlo simulations were made using {\sc
  geant3}~\cite{Geant3}, to design, optimize, and determine the
operating conditions of the spectrometer~\cite{VanHove00}.
All three spectrometer magnets 
have cylindrical symmetry and generate solenoidal fields.
The field intensities, on axis at the center of the magnets, were
1.86, 2.66, and 0.80~T, respectively, for the ``Filter'', ``Tracker''
and ``Lens''.
 
The Filter magnet selects positrons with energies larger than 44~MeV,
emitted into a cone defined by $163^\circ <\theta< 177^\circ$ relative
to the magnet axis, which also defined the average muon polarization axis.
The magnetic field was produced by a split-pair
superconducting coil.
The warm bore was filled with copper scrapers and collimators, shaped
so that only energetic positrons emitted into the above given angles
could pass through.
These obstacles also stop the remaining 28.5 MeV/c positrons that contaminated
the muon beam.

The magnetic field in the Tracker was provided by an 81~cm long superconducting
coil (including two trim coils) generating a uniform field over a large
volume.
The 1~m long by 20~cm diameter warm bore hosted three planes of
double-sided position-sensitive Si strip detectors (SSD) to measure the
positron momentum.  
Inside the Tracker magnet, decay positrons make at least one full turn
of their helix-shaped trajectories.
The positron momentum is determined from the intersections of the
tracks with the three planes of SSDs.

\begin{figure}[!htb]
\centerline{\includegraphics[width=0.75\linewidth]{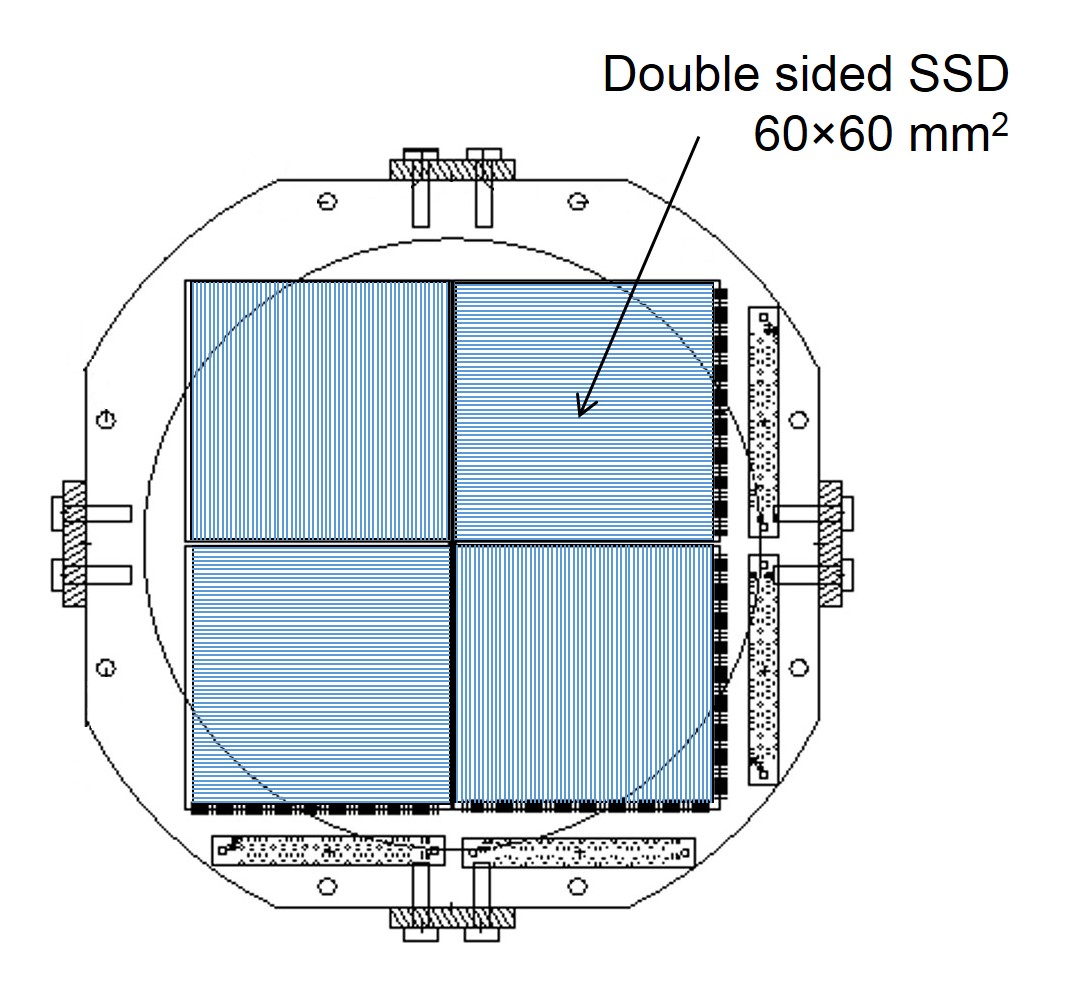}}
\vspace*{-2 mm}
\caption{(Color online) Front view of the arrangement of four SSDs on the
  frame of one of the three planes of detectors located inside the
  Tracker (Fig.~\ref{fig:spectro}). The boards with the front-end electronics
  were connected perpendicular to these planes.}
\label{fig:ssds}
\end{figure}

Each plane contains four $300~\mu\mathrm{m}$ thick
($60\times60~\mbox{mm}^2$) SSDs mounted on an aluminum support as
indicated in Fig.~\ref{fig:ssds}.
Each detector has 60 independent strips per side resulting in a
position resolution of 1~mm in both horizontal and vertical
directions.

The distance between the first and second planes was $z_{21} =
21$~cm and the distance between the second and the third planes was
42~cm.
With this geometry, the radial and longitudinal components of
the momentum are given by~\cite{VanHove00}
\begin{equation}
p_r = \frac{e B d^2_{21}}{\sqrt{4d^2_{21} - d^2_{32}}}\,.
\label{eq:pr}
\end{equation}
and
\begin{equation}
p_z = \frac{ e B z_{21} }{ 2 \arccos{\left[-d_{32}/(2d_{21})\right]}}\,,
\label{eq:pz}
\end{equation}
where $e$ is the electric charge of the positron, $B = 2.66$~T is the
magnetic field intensity and $d_{ij} = \sqrt{(x_j - x_i)^2 + (y_j -
 y_i)^2}$ is the projection on the vertical $(x,y)$ plane of the distance
between hits in the SSD planes $i$ and $j$.

%{\blue The momentum resolution of the spectrometer was smaller than 2.2\%
%FWHM for positrons with energies larger than 47~MeV\@.
%}

Monte-Carlo simulations indicated that the energy resolution of the spectrometer
is 1.15(4) MeV FWHM over the selected energy range. The fits to real data are
fully consistent with this value.
Details about the design, the electronics, and performance of the
spectrometer can be found in Ref.~\cite{VanHove00}.
In particular, both positron transmission rates and momentum
distribution shapes downstream from the spectrometer have been checked
during
preliminary tests and were found to be consistent with the Monte-Carlo
simulations~\cite{VanHove00}.

Figure~\ref{fig:pSpec} shows two energy spectra, after software cuts
(see below), for positrons emitted from the Al and S targets.
%and for
%annihilation-in-flight (AIF) and Bhabha scattering (BHA) events.
%
For comparison purposes the spectra have been
normalized to their respective maxima.
The reduced event rate from the Al target, due to the
muon polarization being maintained, is clearly visible at higher energies.
The points show experimental data and the lines are calculated distributions
including the spectrometer transmission function.

\begin{figure}[!ht]
\center{
\includegraphics[width=\linewidth]{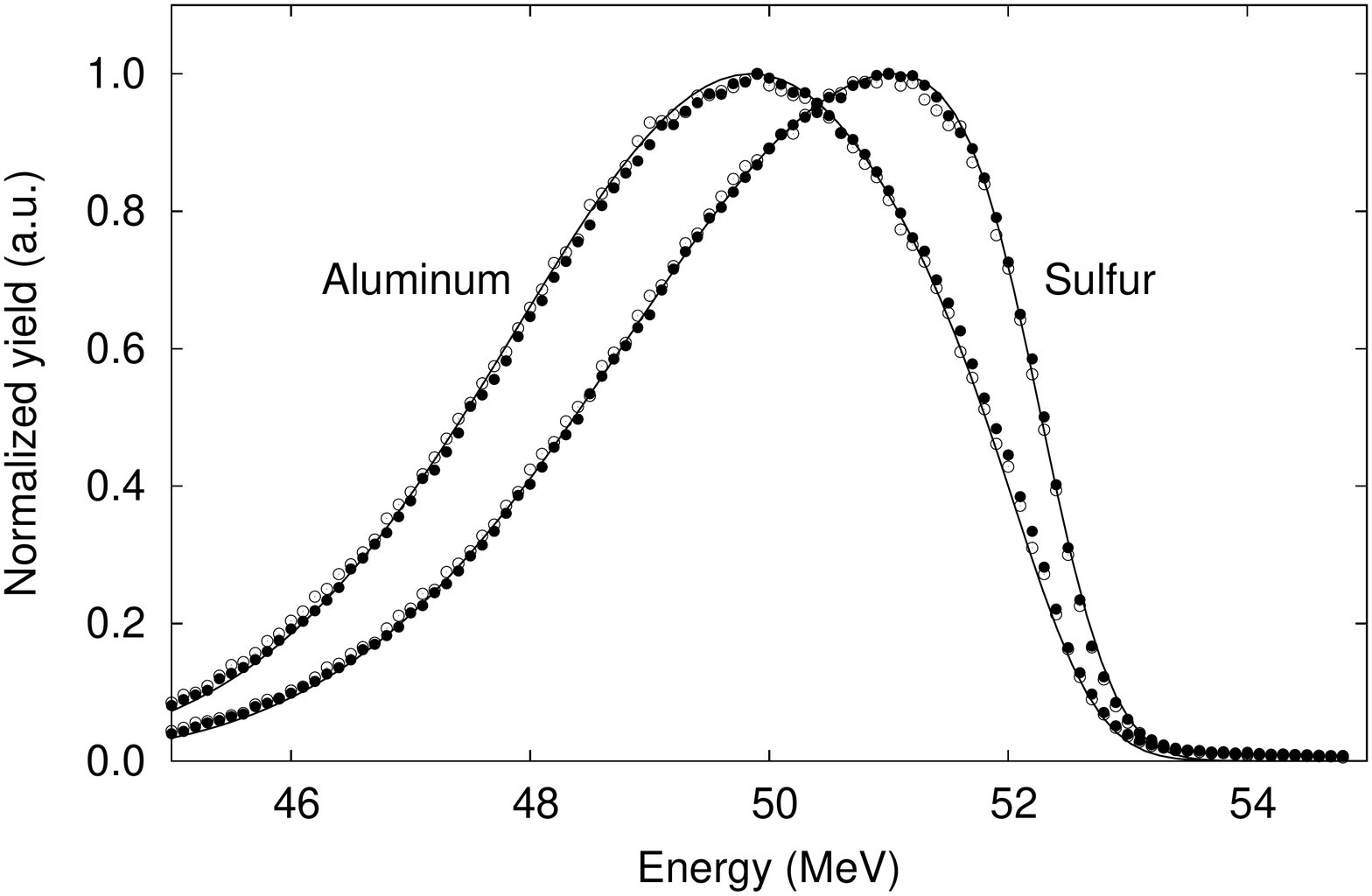}
}
\vspace*{-5mm}
\caption{Typical normalized positron energy spectra as deduced from the
momentum measurement with the SSD located
inside the Tracker magnet for positrons transmitted through the
  spectrometer. The distributions correspond to muons implanted
in the Al target (left curve) and S target (right curve). The full
  circles are for positrons which underwent Bhabha scattering in the
polarimeter and the open circles
  for positrons which annihilated in flight. The solid lines show
  calculated shapes assuming muon polarizations $P_\mu^{\mathrm{Al}} = 0.937$
  and $P_\mu^{\mathrm{S}} = 0.382$.}
\label{fig:pSpec}
\end{figure}

Finally, the Lens magnet guides transmitted positrons into the polarimeter
so that the positron trajectories become essentially
parallel to the spectrometer axis at the location of the magnetized
foils used for the polarization analysis.

%%%%%%%%%%%%%%%%%%%%%%%%%%%%%%%%%%%%%%%%%%%%%%%%%%%%%%%%%%%%%%%%%%%%%%%%%%%%
\subsection{Polarimeter}

The determination of the positron longitudinal polarization was
made using the spin dependence of Bhabha scattering (BHA), 
$e^+ + e^- \rightarrow e^+ + e^-$~\cite{Bincer57,Ford57}, 
and annihilation in flight (AIF), 
$e^+ + e^- \rightarrow \gamma +
\gamma$,~\cite{ApANI,Page57,Fetscher:2007:AIF} processes. 
Data for both processes can be recorded simultaneously, due to the
similar kinematics which, in principle, offers an additional check on
possible systematics since the two processes have analyzing powers with
opposite signs~\cite{Corriveau81}.

\begin{figure}[!ht]
\centerline{
\includegraphics[width=0.9\linewidth]{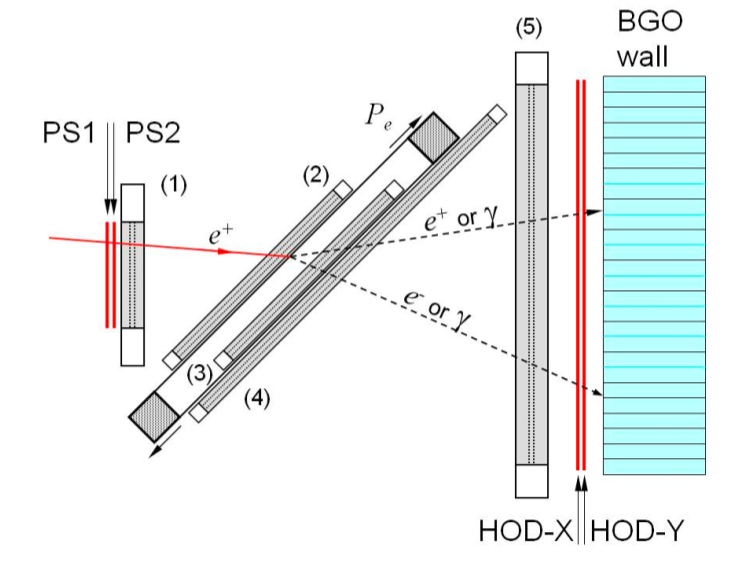}
\vspace*{-2mm}
}
\caption{(Color online) Top view of the positron polarimeter with
  tracings of a scattering event occurring in the first
  magnetized foil of the Vacoflux foil loop.  The plastic
  scintillators (in red) PS1 and PS2 detect incoming positrons, while
  the hodoscope scintillators HOD-X and HOD-Y in coincidence with the
  segmented BGO determine the scattering type.  The MWPC are labeled
  (1)--(5). The Vacoflux foil loop mounted around the yoke provides
  polarized electrons ($P_{e}$).}
\label{fig:polar}
\end{figure}

Incoming positrons were detected at the entrance of the polarimeter
(Fig.~\ref{fig:polar}) by a coincidence between two plastic
scintillators noted PS1 and PS2.
The position at the entrance was determined with the first out of five
multi-wire proportional chambers, MWPC(1), located behind the
plastic scintillators.

Two Vacoflux-50 foils ($75\times15~\mathrm{cm}^2$) mounted on
a loop around a magnetized ARMCO alloy yoke produced electrons with a
polarization $P_e$ oriented in the plane of the foils. Due to the
loop where the foils are mounted, the polarization on the two foils
have opposite directions.
During operation, the foils were first magnetized up to saturation
by two coils wound around the yoke and then left magnetized at their remnant
fields.

For their construction, the foils and the yokes were heated to
$820^\circ$C for about 6~hours in an N$_2$ atmosphere and then slowly
cooled during 10 hours in the presence of an 
external field of 38 Gauss \cite{HeatAnneal}. 
%field
%
% a magnetic field    (B) is in T
% a magnetizing field (H) is in A/m
%
Such treatment generates a sharp hysteresis curve~\cite{Morelle02}
allowing the measurement during the main experiment to be performed
without current in the coils following foil magnetization.
 
The induced magnetic field over the $36\times15~\mathrm{cm}^2$ active
surfaces was 1.910(5)~T\@.
The foil thickness over the active regions was optimized to 0.75(1)~mm
following detailed Monte-Carlo simulations~\cite{Morelle02}.
From the gyromagnetic factor of the alloy and the foil magnetization,
the electron polarization along the foil direction was estimated to be
$P_{e} = 6.88(5)$\%.
Due to the foil orientation by 45$^\circ$ relative to the beam axis, the effective
electron polarization along the spectrometer axis was then
$P_{\mathrm{eff}} = 4.86(3)$\%.
As explained in Sec.~\ref{sec:fits}, it is not necessary to accurately know the
value of the effective electron polarization. 
The foils were sandwiched between three wire chambers, MWPC(2)--(4),
used to locate the vertex where the AIF or BHA scattering occurred.
The two foils, their support yoke, as well as the three wire chambers,
can be rotated as a unit about the vertical axis allowing the
orientation to be changed by $\pm45^\circ$ relative to the beam axis.

\begin{figure}[!ht]
\centerline{
\includegraphics[width=\linewidth]{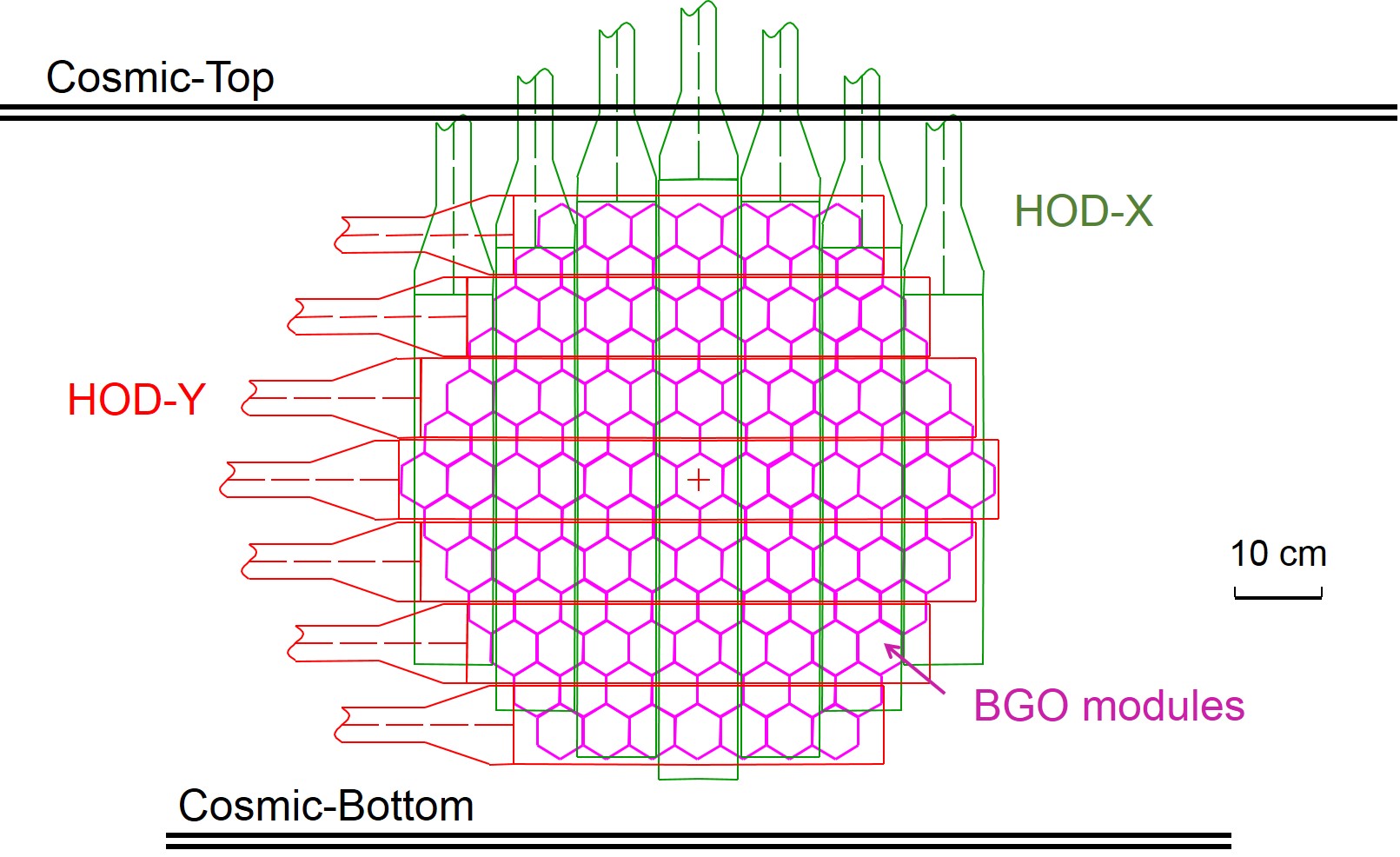}
}
\caption{(Color online) Front view of the hodoscope and the
  calorimeter showing the position of the
  scintillator detectors: HOD-X (in green), HOD-Y (in red) and top
  and bottom cosmic rays telescopes (in black) relative
  to the hexagonal front face of the BGO modules
  of the calorimeter (in magenta).}
\label{fig:hodo}
\end{figure}

The polarimeter is completed by a fifth MWPC, a hodoscope (HOD--X and
HOD--Y) and a calorimeter.
The hodoscope consists of two planes of plastic scintillators, each
having seven slices (90~mm wide and 3~mm thick) of variable lengths
such as to cover the hexagonal front face of the calorimeter (Fig.~\ref{fig:hodo}).
Each panel of the hodoscope was read by a single photomultiplier.

The calorimeter wall consists of 127 BGO crystals, 20~cm long, of
hexagonal section.
This set of detectors was used in the same geometry as in a previous
measurement of the transverse polarization of positrons emitted from
polarized muons~\cite{Danneberg05}.
The BGO wall was surrounded by a thermal shield to stabilize the inner
temperature within $\pm 2^\circ\mathrm{C}$.
In order to limit temperature variations, the high voltage dividers
for the BGO photomultipliers were located outside the thermal shield.
Details about the geometry, temperature stabilization,
photomultipliers, readout electronics and performance of the
calorimeter can be found in Ref.~\cite{Barnett00}.
Two pairs of plastic scintillators panels
(Fig.~\ref{fig:hodo}), $1600\times 160\times 12~\mathrm{mm}^3$ in size
for the top pair and $1220\times 160\times 12~\mathrm{mm}^3$ for the
bottom pair, were located above and below the BGO wall to detect cosmic
rays.
Each panel for cosmic rays detection was read with two photomultiplier tubes,
one at each end of the scintillator.

The entire polarimeter was located inside a
$3\times3\times1.8~\mathrm{m}^3$ iron box, partly indicated on
Fig.~\ref{fig:spectro}, to shield the analyzing Vacoflux foils and the
(otherwise unshielded) calorimeter photomultipliers from the stray
field generated by the spectrometer magnets.
The residual field intensity inside the iron box was smaller than 0.4~mT\@.

\subsection{Signal treatment and data acquisition}

% this combined section unifies the previous two sections called
%
%   \subsection{Triggers and measuring sequences}
%   \subsection{Readout Electronics and Slow Control}
% there is no win in separating them.

Due to the different time responses of the various detectors,
three types of signal readout systems were used: one for the plastic
scintillators (target monitor telescopes, polarimeter entrance
scintillators, hodoscopes, and cosmic muon detectors), another for the
SSD and MWPC signals, and a third for the BGO modules.
The fast plastic scintillator signals were treated with standard
discriminators and coincidence units to construct the logic signals
used in the subsequent triggering decisions.

The BGO detector readout used a set of Lecroy Research Systems
(LRS) Fera modules, LRS-4300B analog-to-digital converters for the
energy information and LRS-3377 time-to-digital converters for the
time information.
The gain stability of the BGO crystals was controlled with independent
LED pulsers located on each module and producing pulses of three
different amplitudes.
The LEDs were triggered by an external clock at 10~Hz.

The SSD and MWPC signals were read using VA-Rich chips (from IDE-AS).
A single chip can read and store signals from 64 channels so that
one chip was used to read and store the 60 channels (one side) of each
silicon detector. To collect the analog data of all $3\times4$
double-side strip detectors, 24 VA-Rich were used.
When triggered, a V551B CAEN VME module drove all VA-Rich chips in parallel,
first with the hold signal to lock in the event of interest, then with
the multiplexer signals to read the data into V550 CAEN modules
where the signals were digitized and noise subtracted.
The data was then transferred to the acquisition computer.
A~similar system with 15 VA-Rich chips was used for the MWPC readout.
The full reading time per event took roughly $700~\mu\mathrm{s}$.

After shaping and discrimination, the logic indicating any of several
different events (detailed below) was realized with ALTERA
programmable logic gate arrays.
All plastic scintillators as well as the combined BGO signals
entered the trigger logic whereas data from the SSD and the MWPC were
read only when required, based on the trigger type.
To reduce pile-up events which could lead to the misidentification
between the signals of an incoming positron and those of outgoing
particles, an updated dead time of $3~\mu\mathrm{s}$ was imposed on
the trigger logic once a positron was detected at the entrance of the
polarimeter.
The fast acquisition system was based on a vxWorks front-end hosted in
CAMAC and VME combined with back-end software developed by the group at
Louvain--la--Neuve.

A slow control system, based on LabView, was used to set and monitor
other parameters of the apparatus such as: {\it i)} the foil
magnetization current and the sequence for reversing the foil
magnetization; {\it ii)} the high voltage of the plastic scintillators
and BGO detectors; {\it iii)} the number and the amplitude of each of
the LED signals sent to the BGO modules; and {\it iv)} the measurement
and control of the temperature inside and outside the BGO thermal
shield.
The current for the loop magnetization was reversed by the slow
control system after every one-hour run.
Data with the depolarizing S~target was taken for 12~hours after every
two days of measurement with the polarization-maintaining 
 Al  target.
The magnetized foils together with MWPC(2)--(4) were rotated
between $+45^{\circ}$ and $-45^{\circ}$ every four days.

The data acquisition included eight mutually exclusive triggers
running simultaneously. Table~\ref{tab:triggers} gives their names and
comparative rates for polarized and unpolarized muons.
The trigger for the primary AIF and BHA events, or for noninteracting Michel
(MIC) positrons was generated
by an incoming positron by combining three signals: 
1) the coincidence (PS1$\cap$PS2) providing the time-zero
reference for the event;
2) the signals from HOD--X, HOD--Y, and MWPC(5) which entered the
trigger decision via a hardware selection of the 
multiplicity of the detectors: 0 for events identified as
AIF, 2 for BHA, and 1 for the most frequent MIC events 
arising from positrons simply crossing the polarimeter; and,
3) a fast summed amplitude signal from the BGO giving
the total energy deposited in the calorimeter, with a threshold set at
30~MeV\@.
The MIC triggers were prescaled by a factor 50 to reduce the load on
the data acquisition system.
LED signals and cosmic muons crossings the BGOs were additional
triggers.
The triggers from the plastic scintillator telescopes located around
the muon implantation target,
%recorded in lifetime mode,
were independent from those of events in the polarimeter
and were used for on-line monitoring of the muon beam intensity and
polarization.

%%%%%%%%%%%%%%%%%%%%%
\begin{table}[t]
\caption{Average event rates (in $\mathrm{s}^{-1}$) for each kind
  of trigger at a typical primary proton beam current of 1.6~mA\@.
  The bottom line indicates the corresponding acquisition dead-time,
  which was mostly due to the conversion and reading time of the
  V550--V551B modules.}
\label{tab:triggers}
\begin{center}
\begin{tabular}{l@{\hspace{3mm}}r@{\hspace{5mm}}r}
\hline \hline
Trigger source & Al target &  S target \\ \hline
Annihilation (AIF)         & 64  & 160 \\
Bhabha       (BHA)         & 79  & 205 \\
Michel/50    (MIC)         & 47  & 122 \\
Cosmic                     &  0.5&  0.5\\
LED                        &  8  &   8 \\
Telescope($121^\circ$)/100 & 19  &  16 \\
Telescope($+90^\circ$)/100 & 12  &  12 \\
Telescope($-90^\circ$)/100 & 14  &  14 \\
\hline
Dead Time                  &15\% & 34\%\\
\hline \hline
\end{tabular}
\end{center}
\end{table}
%%%%%%%%%%%%%%%%%%%%%

%%%%%%%%%%%%%%%%%%%%%%%%%%%%%%%%%%%%%%%%%%%%%%%%%%%%%%%%%%%%%%%%%%%%%%%%%%%%
\section{Data Analysis}

A total of 501 data sets, each of about one hour duration,
have been collected during the experiment.
From this set, numerous runs correspond to tests made at the beginning
and at the end of the experiment.
From the remaining files, 255 runs were selected by the filters for
which the proton beam intensity was stable during the measurement and all
components of the apparatus and electronics operated without fault.
The data analysis was applied to this final set of files which contained
comparable statistics for the different configurations.

%%%%%%%%%%%%%%%%%%%%%%%%%%%%%%%%%%%%%%%%%%%%%%%%%%%%%%%%%%%%%%%%%%%%%%%%%%%%
\subsection{Calibration of the BGO modules}

The energy calibration of the BGO modules was performed using both MIC and
cosmic events.
The momentum of a MIC event is first measured by the SSD tracker following
Eqs.~(\ref{eq:pr}) and~(\ref{eq:pz}).
The energy of the same MIC positron as measured in the calorimeter results
from its convolution with the polarimeter transmission function.
That function is determined via Monte-Carlo simulation and includes
the energy losses in the plastic scintillators, in the MWPCs, and in
the Vacoflux foils.
 
Cosmic ray muons deposit a rather constant amount of energy in each
module, and the trajectory of a cosmic event can be easily
reconstructed when five or more BGO modules are hit.
With fewer involved modules, the determination of the energy loss per
length becomes difficult, in particular in the outer region of the
wall.
Because of that difficulty, the BGO modules were calibrated only once
using a large set of cosmic ray events collected during a period when
there were no muons from the beam.
 
The module gain stability was monitored with
LED signals of three different amplitudes.
The three LED peak positions were observed to drift by 
up to 10\% relative amplitude over the 25 day duration of the effective measurement.
Such drift was not caused by variations in the LED light output since
correlated drifts were observed for the MIC spectra centroid as
seen by the BGO wall.
By deconvoluting the measured distribution from the polarimeter
transmission function, the average energy resolution of the BGO wall
%
%was found to be $\sigma_{BGO}$ = 4.4~MeV (standard deviation) at
%42~MeV\@.
%
% (* 2.0 (sqrt (* 2.0 (log 2.0))) 4.4 MeV) = 10.36 MeV
was found to be 10.4~MeV FWHM at 42~MeV.

%%%%%%%%%%%%%%%%%%%%%%%%%%%%%%%%%%%%%%%%%%%%%%%%%%%%%%%%%%%%%%%%%%%%%%%%%%%%
\subsection{Tracking efficiencies}

The efficiencies of the MWPC and the plastic-scintillator hodoscope
have also been determined using MIC events.
Reconstructed tracks for ``perfect'' MIC events contain only
one hit in each plane of each MWPC as well as a single hit 
on the plastic scintillator hodoscope, with one signal on the
vertical and one on the horizontal directions.
The detector inefficiencies are found by comparing the rates of such
``perfect'' Michel events to those where one of the expected hits is
missing.

Among the ten MWPC wire planes, the smallest efficiencies were 
consequently found
to be 96.1\% for the plane giving the horizontal position in MWPC(4)
and 97.3\% for the plane giving the vertical position in MWPC(3).
All other planes had efficiencies larger than 99.0\%.

Because the hodoscope scintillators do not overlap
each other, the surface covered by the scintillators has long thin
gaps both horizontally and vertically (Fig.~\ref{fig:hodo}).
The hodoscope efficiency was 95.9\% horizontally and 98.7\%
vertically, roughly in accordance with the $~2$--$3$~mm separation
between the 90~mm wide scintillator strips.
This separation was
caused by the individual scintillator light-tight wrapping.
By extrapolating the MWPC track information to the hodoscopes for
tracks with missing hodoscope hits, it was clearly shown that the
missing hits corresponded to the small gaps between adjacent scintillators.

%%%%%%%%%%%%%%%%%%%%%%%%%%%%%%%%%%%%%%%%%%%%%%%%%%%%%%%%%%%%%%%%%%%%%%%%%%%%
\subsection{Data selection}

\begin{figure}[hb]
\vspace*{-3 mm}
\centerline{\includegraphics[width=\linewidth]{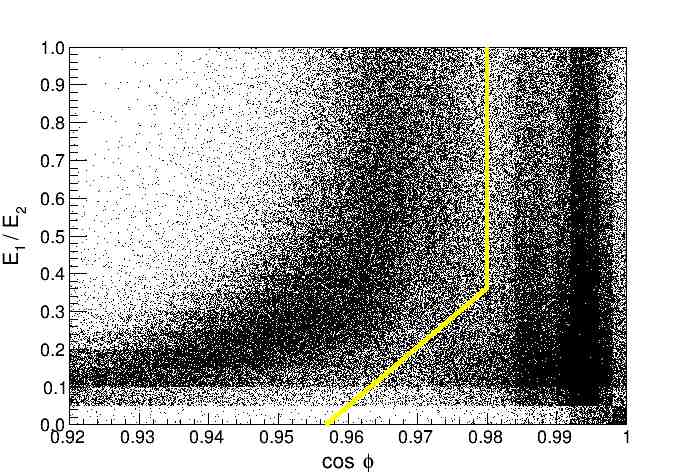}}
\caption{(Color online) Two-dimensional distribution of events selected by the AIF
  trigger plotted as a function of the
  measured $\cos{\phi}$, where $\phi$ is the angle between the two photons
  in the laboratory system and of the ratio $E_1/E_2$
  between the smallest and the largest energies of the photons.
  The sweeping distribution at the left corresponds to AIF events
  following Eq.~(\ref{eq:kineAIF}). The line shows the cut limit applied
  to both AIF and BHA events.}
\label{fig:kinematics}
\end{figure}

The AIF and BHA events have a two-body final state following
the reactions in one of the Vacoflux foils.
In the laboratory frame, the opening angle, $\phi$, between the
two outgoing particles can be
determined from the event topology, using the position
information from the MWPCs. The relation between this angle and
the total energies, $E_1$ and $E_2$,
of the outgoing particles having both a rest mass $m$, is
obtained from simple kinematics
\begin{equation}
\cos{\phi} =\frac
{ E_1 E_2+ m^2 c^4 -  m_{e} c^2(E_1+E_2) }
{\sqrt{E_1^2-m^2c^4} \sqrt{E_2^2-m^2c^4} }
\label{eq:kineBB}.
\end{equation}
For BHA events, $m = m_e$, where $m_e$ is the electron mass,
whereas for AIF events ($m = 0$) this equation simplifies to
\begin{equation}
\cos{\phi} = 1 - m_e c^2 \frac{E_1 + E_2}{E_1 E_2}.
\label{eq:kineAIF}
\end{equation}
It is convenient to use
Eqs.~(\ref{eq:kineBB}) and (\ref{eq:kineAIF}) for a kinematic identification
of the signal for the scattered events in a
two dimensional histogram plotting the ratio, $E_1/E_2$, between the
smallest and the largest of the two energies versus $\cos{\phi}$.
Such signature was clearly visible for AIF events
(Fig.~\ref{fig:kinematics}) but not for BHA events, possibly due to
scattering in matter between the foils and the calorimeter and to the
contribution of misidentified background.

Within the energy interval selected in the experiment, BHA events
following Eq.~(\ref{eq:kineBB}) are expected to have a very similar signature
to the one observed for AIF events in the two-dimensional distribution of
Fig.~\ref{fig:kinematics},
since the electron rest mass is small compared to the total energies
of the outgoing electron and positron.
Consequently, 
for the uniformity of treatment, 
the same kinematic cut  
was applied to BHA and AIF data as indicated 
by the line in Fig.~\ref{fig:kinematics},
to separate the signal from background events visible at small angles.
Additionally, an energy cut required both $E_1$ and $E_2$ to be each
larger than $6$~MeV, and an upper limit of
$(E_1+E_2) < 70$~MeV was also imposed to reject accidentally summed events.

After passing checks of calorimeter gain stability and calibration,
the off-line analysis proceeded in three steps:
{\it i)} the positron momentum is determined from the hits of the SSD
located inside the Tracker magnet;
{\it ii)} evaluations are made of the energies and barycenter
positions of the clusters created by the two reaction products as
detected by the calorimeter;
and {\it iii)} the event vertex and scattering foil is reconstructed
from the MWPCs data.
No background subtraction has been performed on the selected events.

\subsection{Super-ratios}

For each scattering process (BHA and AIF) the events were sorted into
one of $2^4$ different types according to the experimental conditions
of the four main measurement parameters:
{\it i)} the foil where the scattering occurred (1st or 2nd);
{\it ii)} the current polarity for the Vacoflux loop magnetization
($\pm$: positive or negative); 
{\it iii)} the magnetized foil orientation relative to the beam axis
($\pm45^\circ$); and
{\it iv)} the implantation target (Al for polarization maintaining, or
S for depolarizing).

For a given run, the number of events originating from the 1st
and 2nd foils, $y_{1}$ and $y_{2}$, are measured simultaneously. It is
then
convenient to take the ratio of those numbers to avoid the use of an
external normalization.
These ratios are the primary information extracted from each run and
are noted $r^\pm_\alpha = y_{1,\alpha}^\pm/y_{2,\alpha}^\pm$, where
$\pm$ indicates the magnetization-current polarity and $\alpha$ stands
for the eight remaining experimental conditions.

Under magnetization current-polarity reversal, effects associated with
the electron polarization do change sign but effects from detector
geometry do not.
From the ratios $r^\pm_\alpha$ introduced above one then defines
the ``super-ratios'' $s_\alpha$ by
\begin{equation}
s_\alpha =
\frac{r^+_\alpha - r^-_\alpha}{r^+_\alpha + r^-_\alpha}\,.
\label{eq:superrat}
\end{equation}

%In order to estimate the maximal possible super-ratios, it is
%convenient to introduce the solid angles for each configuration,
%$\omega_{i,\alpha}^{\pm}$.
%
As will be shown in Sec.~\ref{sec:analyzepower}, differences in solid
angles $\omega_{i,\alpha}^{\pm}$ from foils $i=1$ and 2 relative to the
polarimeter detectors as well as effects due to different incident positron
intensities on the foils cancel in the super-ratios under the
assumption that 
%\begin{equation}
$ \omega_{1,\alpha}^+ \omega_{2,\alpha}^-  = \omega_{1,\alpha}^- \omega_{2,\alpha}^+$.
%\end{equation}
%
%which holds reasonably well.

Since the data are finally analyzed as a function of the energy
obtained from the
positron momentum as measured by the SSD Tracker, the super-ratios
are sorted into 20 energy bins from 45 to 55~MeV\@.
%.
This energy sorting results in rather small statistics per bin
within each run. Moreover the combination of selected runs having opposite
polarization currents to form the super-ratios could possibly induce a bias.
Consequently,
the statistics from all the 255 runs has been grouped and sorted
for the different configurations.
This generates $2^5$ summed event vectors of 20 energy bins,
corresponding
to the different running conditions (foil number, current
polarity for the magnetization, orientation of the magnetized foil,
implantation target Al or S) and event type (AIF or BHA). 

\begin{figure}[t]
\centerline{\includegraphics[width=\linewidth]{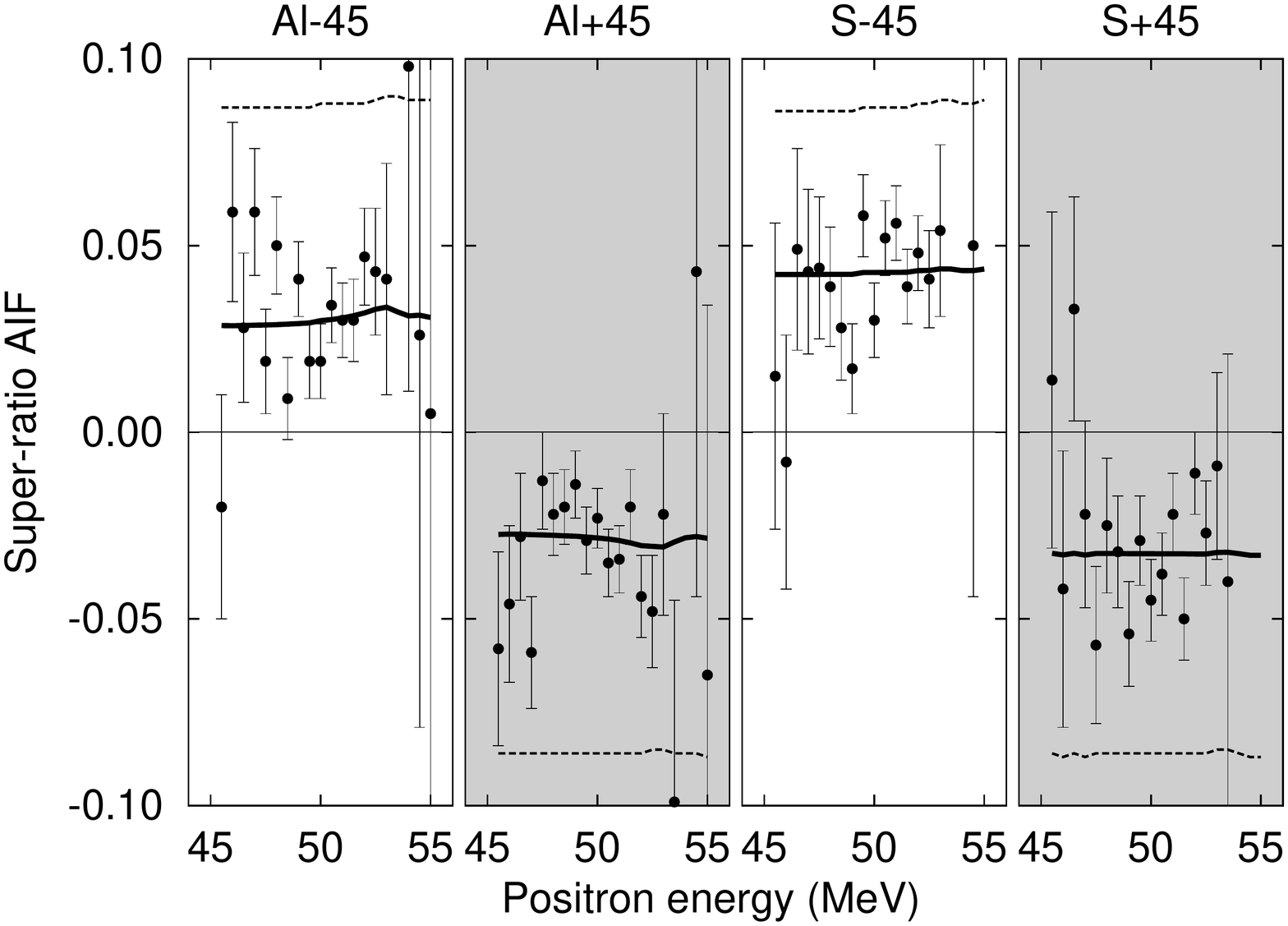}}
\centerline{\includegraphics[width=\linewidth]{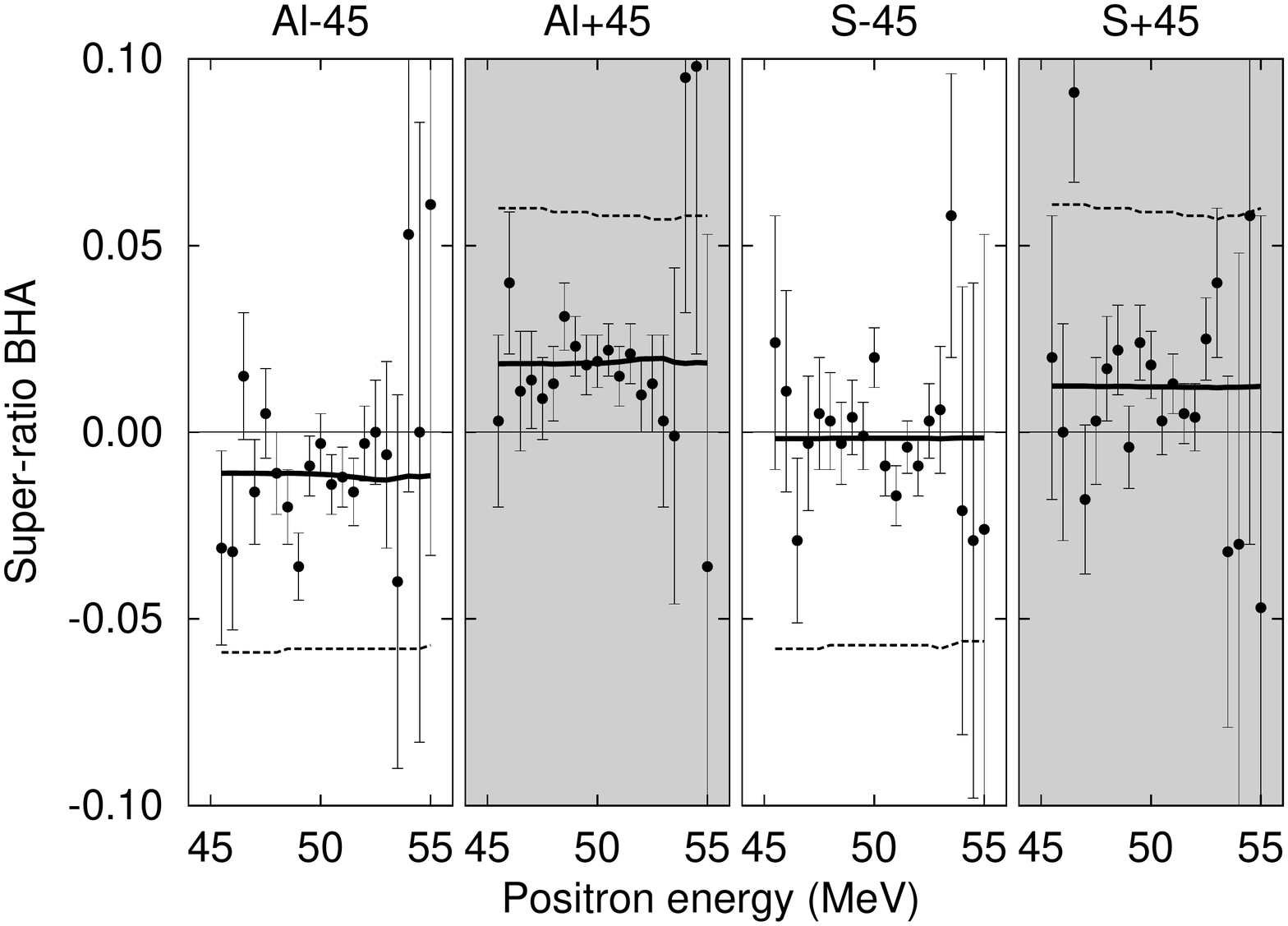}}
%vspace*{60mm}
\caption{Experimental super-ratios, Eq.~(\ref{eq:superrat}),
  for the two targets and the two
  orientations of the scattering foils as a
  function of the positron energy for AIF (upper panel) and BHA (lower
  panel). The dotted lines show the values of the function
  $\varphi(A^\pm_{i,\alpha})$ which is the super-ratio calculated from
  the analyzing power assuming a positron polarization $P_L
  = 1$ (Sec.~\ref{sec:longpol}). The solid lines correspond to the fit of the
  super-ratios (Sec.~\ref{sec:fits}).}
\label{fig:superrat}
\end{figure}

The eight sets of super-ratios calculated for each of the
two processes, the two target types, and the two orientations of the
magnetized foil are shown in Fig.~\ref{fig:superrat} as a function of
the positron energy.
The sign inversion of the asymmetries
under the $\pm45^\circ$ rotations of the magnetized foils is clearly
visible.
The asymmetries for AIF are seen to be larger than for BHA and with
opposite signs as expected from the analyzing powers of these
processes~\cite{Corriveau81}.
%

%%%%%%%%%%%%%%%%%%%%%%%%%%%%%%%%%%%%%%%%%%%%%%%%%%%%%%%%%%%%%%%%%%%%%%%%%%%%
\subsection{Analyzing powers}
\label{sec:analyzepower}

In order to estimate the maximal values expected for the super-ratios
and to study
the energy dependence of the analyzing powers,
the measured ratios of events between foil~1 and~2 can be expressed as
a function of solid angles, the longitudinal polarization, and analyzing powers:
\begin{equation}
r^\pm_\alpha = 
\frac{\omega^\pm_{1,\alpha} (1 + P_L A^\pm_{1,\alpha})} 
       {\omega^\pm_{2,\alpha}(1 + P_L A^\pm_{2,\alpha})}
%, \hspace{3mm}
%r^-_\alpha = 
%\frac{\omega^-_{1,\alpha}(1-P_L A^-_{1,\alpha})} 
%{\omega^-_{2,\alpha}(1-P_L A^-_{2,\alpha})},
\label{eq:superrat1}
\end{equation}
where $A^{\pm}_{i,\alpha}$ are the analyzing powers for each
configuration and type of process and include the relative sign
associated with the selected geometry and scattering process.
Again, subscript $i=(1,2)$ refers to the Vacoflux foil in which the
scattering occurred, and $\alpha$ indexes the other
eight experimental
conditions associated with the foil orientation, the two scattering processes
and the implantation target.

The analyzing powers, $A^{\pm}_{i,\alpha}$, were extracted by using the
kinematic variables from experimental data and the cross sections
of the scattering processes as calculated from quantum electrodynamics (QED).
For each scattering event, $l$, the incoming positron energy 
and the kinematics of the outgoing particles are used to compute the
corresponding raw analyzing power, $(a_p)_l$, of the associated process
(BHA~\cite{Ford57}, AIF~\cite{Page57,ApANI}), assuming  $P_L=1$.
The MWPC tracking data are used to determine
the angle, $\theta_{s}$, between the direction of the electron spin
in the struck Vacoflux foil
and that of the momentum of the incoming positron, thus providing a
weight factor, $(\cos{\theta_s})_l$, for $(a_p)_l$.
Table~\ref{tbl:Appol} gives the absolute value of the mean and RMS for the
$a_p$ and $a_p\cos{\theta_s}$ distributions obtained from all events and
configurations for each of the two processes (AIF or BHA).
In practice, $\theta_s$ is very close to $45^\circ$ (the mean is
$44.3^\circ$ for AIF and $44.6^\circ$ for BHA) since the direction
of incidence of the positrons is almost parallel to the spectrometer axis.

\begin{table}[t]
\caption{Absolute mean and standard deviations of the distributions
  of the calculated analyzing powers, $a_p$, and of the projections
  $a_p \cos{\theta_s}$.
%with $\theta_s$ the angle between the incident positron
%spin and the electron spins in the Vacoflux foil.
}
\label{tbl:Appol}
\begin{center}
\begin{tabular}{l@{\hspace{3mm}} r@{\hspace{5mm}} l@{\hspace{5mm}}    
                                 r@{\hspace{5mm}} l@{\hspace{5mm}} 
                                                  c@{\hspace{2mm}}}\hline\hline
         &\multicolumn{2}{c}{$a_p$} &\multicolumn{2}{c}{$a_p\cos{\theta_s}$} & Number of \\
 Process &  Mean &  RMS   & Mean  & RMS & events \\
\hline
AIF &  0.88 & 0.022  & 0.63  &0.066  &$1\,564\,835$\\
BHA &  0.59 & 0.15   & 0.42  &0.11   &$2\,313\,549$  \\
\hline\hline
\end{tabular}
\end{center}
\end{table}%

\begin{figure}[t]
\centerline{\includegraphics[width=\linewidth]{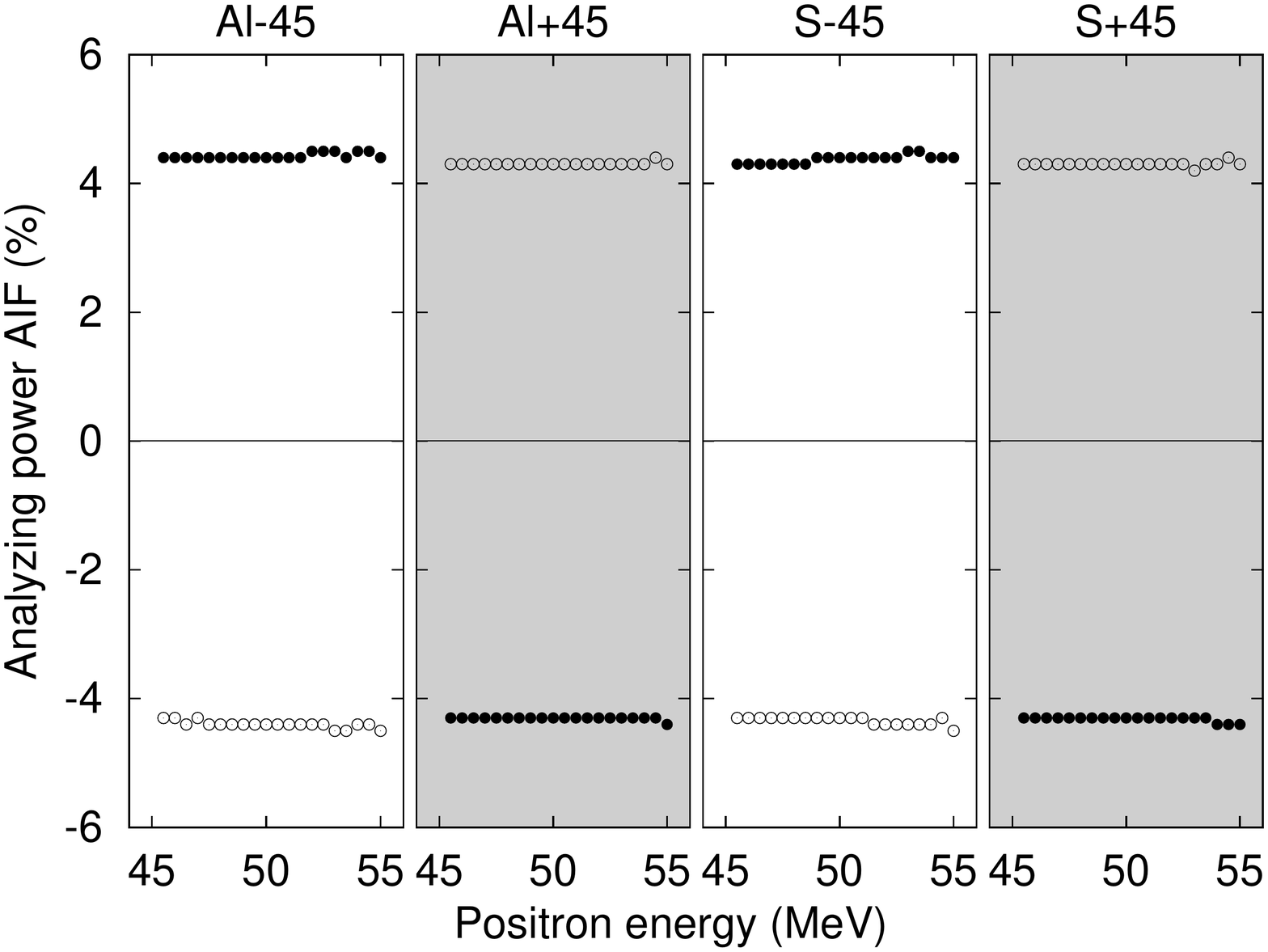}}
\centerline{\includegraphics[width=\linewidth]{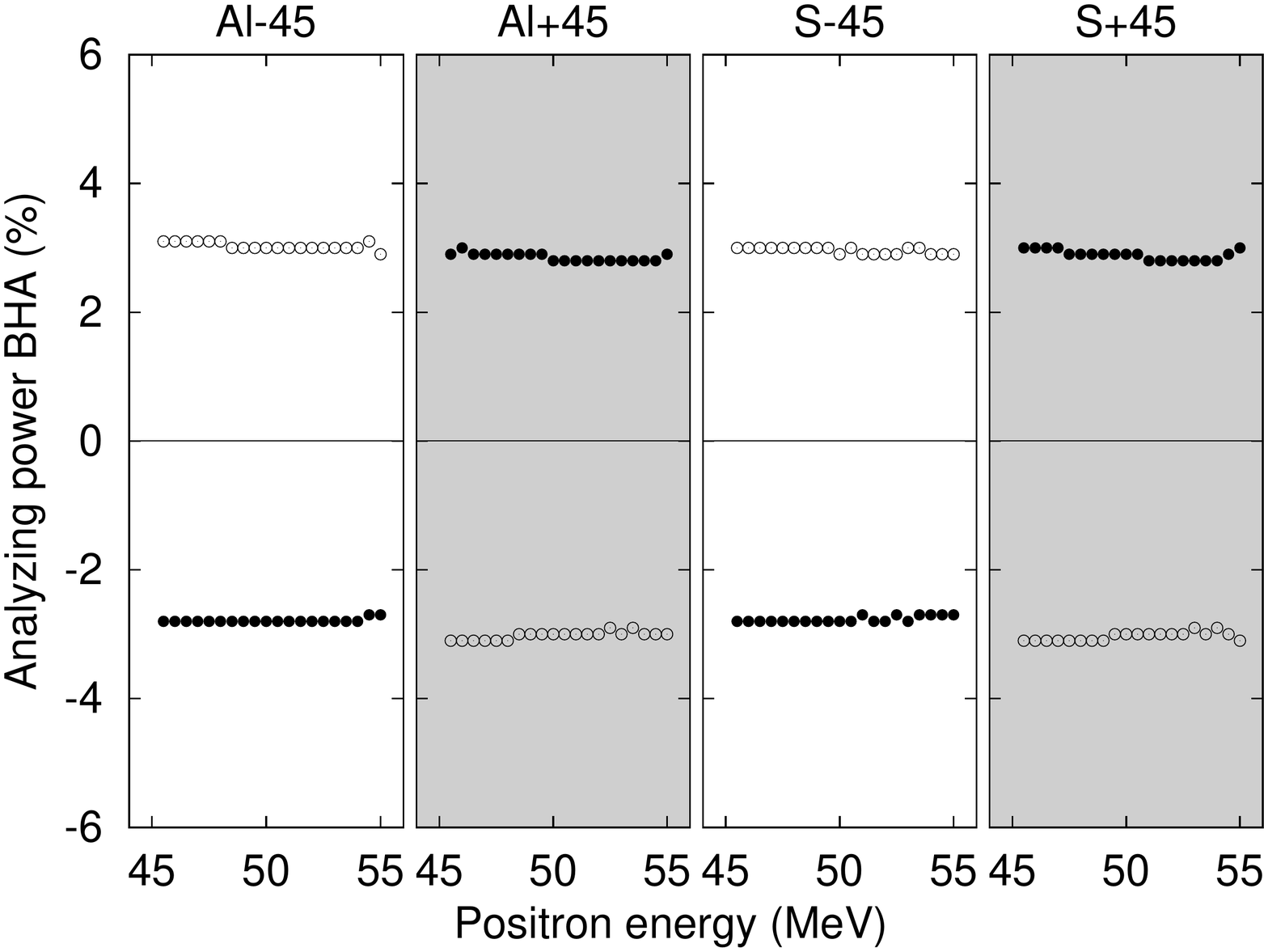}}
%vspace*{60mm}
\caption{Calculated analyzing powers $A^{-}_{i,\alpha}$,
corresponding to the negative magnetization current,
as a function of the positron energy for AIF events (upper panel)
and BHA events (lower panel). The open (closed) circles are for events
from scattering foil 1 (2). 
The values for the positive magnetization current, $A^{+}_{i,\alpha}$,
have a similar trend but with opposite signs and are not shown.}
\label{fig:apn}
\end{figure}

Like for the super-ratios, the analyzing powers and their
projections were sorted into 20 energy bins and further
classified by the $2^4$ experimental conditions for the two scattering
processes. 
%
%8 $\alpha$ configurations and  moreover separated  into 4 sets
%corresponding to the two foils and the two magnetization of the
%Vacoflux foils.   
%
%8 sets for the same configurations than the super-ratios. The result
%are shown in Fig.~\ref{fig:apn}.
%
The energy distributions resulting  from this sorting were then
multiplied by the electron polarization $P_{e}$ in the
Vacoflux foils, and each energy bin was normalized by the number,
$n$, of values used in that bin.
These operations can be summarized by the following expression for
the calculated analyzing power at each energy bin
\begin{equation}
A^{\pm}_{i,\alpha} = 
   \left[ \frac{P_{e}}{n} \sum_{l=1}^n (a_p)_l (\cos{\theta_s})_l\right] 
   ^{\pm}_{i,\alpha}\,.
\end{equation}
The distributions of the analyzing power corresponding to the negative
magnetization current are shown in Fig.~\ref{fig:apn} for AIF and BHA events.
%
%The expected sign variations between foil number, opposite magnetic
%field condition, and scattering process are clearly visible.
%
The distributions associated with the positive
magnetization current are not shown since they behave similarly
except for their global sign reversal compared with the
negative magnetization. 

The main conclusion from this study is that the analyzing powers of the
two processes as calculated within QED are, to a sufficient approximation,
constant over the measured interval of the positron energy.
In particular, the asymmetries do not display any significant variation
towards the end-point energy. 

%%%%%%%%%%%%%%%%%%%%%%%%%%%%%%%%%%%%%%%%%%%%%%%%%%%%%%%%%%%%%%%%%%%%%%%%%%%%
\subsection{Longitudinal polarization}
\label{sec:longpol}

The super-ratios in Eq.~(\ref{eq:superrat}) can be expressed as a
function of the positron longitudinal polarization by using
Eq.~(\ref{eq:superrat1}), and assuming that $\omega_{1,\alpha}^+
\omega_{2,\alpha}^- = \omega_{1,\alpha}^-\omega_{2,\alpha}^+ $:
\begin{eqnarray}
s_\alpha & = &
P_L \frac{f_1(A^{\pm}_{i,\alpha}) +P_L f_2(A^\pm_{i,\alpha})}
         {2{+}P_L f_3(A^{\pm}_{i,\alpha}) {+} P^2_L f_4(A^\pm_{i,\alpha})}
\label{eq:superrat2-1} \\
       & = &P_L \varphi(A^{\pm}_{i,\alpha},P_L)\,.
\label{eq:superrat2-2}
\end{eqnarray}
The functions $f_j(A^{\pm}_{i,\alpha})$ have the following dependences
on the analyzing powers 
\begin{eqnarray}
f_1 &=& (A^+_1 + A^-_2) -  (A^-_1 + A^+_2) \\
f_2 &=& A^+_1  A^-_2 -  A^-_1  A^+_2 \\
f_3 &=& (A^+_1 + A^-_2) + (A^-_1 + A^+_2) \label{eq:f3} \\
f_4 &=& A^+_1  A^-_2 +  A^-_1  A^+_2 
\end{eqnarray}
where the subscripts $\alpha$ were omitted for clarity.
In order to estimate the remaining dependence of the function
$\varphi(A^{\pm}_{i,\alpha},P_L)$ on the positron longitudinal
polarization, the function was computed for
the extreme values $P_L=0$
and $P_L=1$. 
For all values of the reduced energy within the selected energy interval,
this comparison  shows  a variation
 $0.998 \le \varphi(A^{\pm}_{i,\alpha},1) \; / \varphi(A^{\pm}_{i,\alpha},0) \le  0.999$
%$\varphi(A^{\pm}_{i,\alpha},1) / \varphi(A^{\pm}_{i,\alpha},0) \le 1.0022$.
%
which can be neglected at the
current level of precision. The polarization $P_L$ in
$\varphi(A^{\pm}_{i,\alpha},P_L)$ was then fixed to $P_L=1$.
Equation~(\ref{eq:superrat2-2}) then becomes  a linear function of the positron
longitudinal polarization:
\begin{equation}
s_\alpha = P_L \varphi(A^{\pm}_{i,\alpha})\,,
\label{eq:superrat3}
\end{equation}
where the dependence of $\varphi$ on the positron
polarization has been omitted.

Since the analyzing powers $A^+_i$ and $A^-_i$ are of similar magnitude
but have opposite signs we have $f_3 \approx 0$ in Eq.~(\ref{eq:f3}).
Next, since the linear terms in the analyzing powers dominate the super-ratio
given in Eq.~(\ref{eq:superrat2-1}), 
the leading term is given by the function $f_1$ so that, 
within the accuracy of this experiment,
$\varphi(A^{\pm}_{i,\alpha})$ is approximately equal to
$A^{+}_{1,\alpha} + A^{-}_{2,\alpha} \approx 2A^{+}_{1,\alpha}$.

%Figure \ref{fig:apn} shows the calculated analyzing powers as a function
%of the positron energy.
The functions $\varphi(A^{\pm}_{i,\alpha})$ are
superimposed on the data in Fig.~\ref{fig:superrat} for
the eight ${\alpha}$ configurations.
The differences between the average measured super-ratios and the calculated
functions are discussed in Sec.~\ref{sec:fits}.

%%%%%%%%%%%%%%%%%%%%%%%%%%%%%%%%%%%%%%%%%%%%%%%%%%%%%%%%%%%%%%%%%%%%%%%%%%%%
\subsection{Enhancement factors}

For each target type ($T=$Al or S), the positron transmission rate
through the spectrometer varies with the positron energy and
emission angle as
%
%{\red J'utilise ici $n$ au lieu de $y$ car tu as soulign\'e plus
%haut que $y$ indique un nombre d'evenements.}
%
\begin{equation}
n^{T}(x) \propto \int_\Omega \! \left[F(x) + \xi P^{T}_\mu G(x,z) \right]d\Omega \,
\label{eq:yield}
\end{equation}
where $F(x) = x^2(3-2x)$, $G(x,z) = x^2 z (2x-1)$ and the integral is performed
over the angular acceptance of the spectrometer.
The equations for the two targets can be combined to form the ratio,
\begin{equation}
\frac{n^{\mathrm{Al}}(x){-}n^{\mathrm{S}}(x)} {n^{\mathrm{Al}}(x)} = 
\frac{(P^{\mathrm{Al}}_\mu-P^{\mathrm{S}}_\mu) \xi \int_\Omega \; G(x,z) d\Omega}
 {\int_\Omega \; \left[ F(x) + \xi P^{\mathrm{Al}}_\mu \; G(x,z) \right] d\Omega}.
\label{eq:deltayield}
\end{equation}
The factor $x^2$, which enters $F(x)$ and $G(x,z)$, cancels in the 
expression given in Eq.~(\ref{eq:enhfact}) extracted from the double differential
decay rate.
The right-hand side of Eq.~(\ref{eq:deltayield}) contains then the
kinematic factor that appears on the right-hand side of
Eq.~(\ref{eq:enhfact}).
Consequently, the actual enhancement factors can be expressed as
a function of measured quantities, after integrating the rates over
the spectrometer acceptance. For each of the two targets $T$ we have
\begin{equation}
{k( P^{T}_\mu,x)} =
\frac{n^{\mathrm{Al}}(x)- n^{\mathrm{S}}(x)} {n^{T}(x)} 
\frac {P^{T}_\mu}  {(P^{\mathrm{Al}}_\mu-P^{\mathrm{S}}_\mu)}\,.
\label{eq:enhfactT}
\end{equation}
This expression
is composed of two factors: the first is determined from the
experimental yields and is strongly energy dependent; the
second is determined from the muon polarization in the target and
is energy and geometry independent.
The first factor could potentially differ for each of the
$\alpha$--labeled configurations since the geometry of the selected
events can vary.
For that reason the transmitted positrons selected for the determination
of the enhancement factors
are those that undergo either AIF or BHA scattering, chosen
independently of the foil in which the scattering process took place
but otherwise sorted following the eight $\alpha$ configurations
and the two magnetizing currents.

%summed by their energy, and normalized by the telescope data recorded
%under the same $\alpha$ conditions. 

%{\bf ONC: I frankly do not consider that the texts in color below
%clarify how the enhancement factors were calculated!!!} 
%{\blue These events are summed and ordered  in the usual 20 energy
%bins and sorted  in only $2^4$ configurations since the foil number
%doesn't appear here. 
%Same is done for the sum of the two telescopes placed at $90^\circ $
%at the entrance of the spectrometer, loading  each the energy
%channels with the same telescope value but different for each
%configuration. 
%Each event yield vector, is then  normalized  by  its corresponding
%telescope vector content. 
%
%Energy dependent part of equations  (\ref{eq:enhfact2}) and
%(\ref{eq:enhfact3}) are then  applied resulting in   $2^4 \times 20 $
%values which can be divided in 2 groups of the usual $\alpha$
%configurations; one for positive magnetization  and the other for
%negative magnetization. }  
%
%Since the AIF and BHA events are extracted from a common data set
%they are equally normalized. 

%{\blue These two groups are  superimposed in Fig.~\ref{fig:enh}.}

\begin{figure}[!hb]
\centerline{\includegraphics[width=\linewidth]{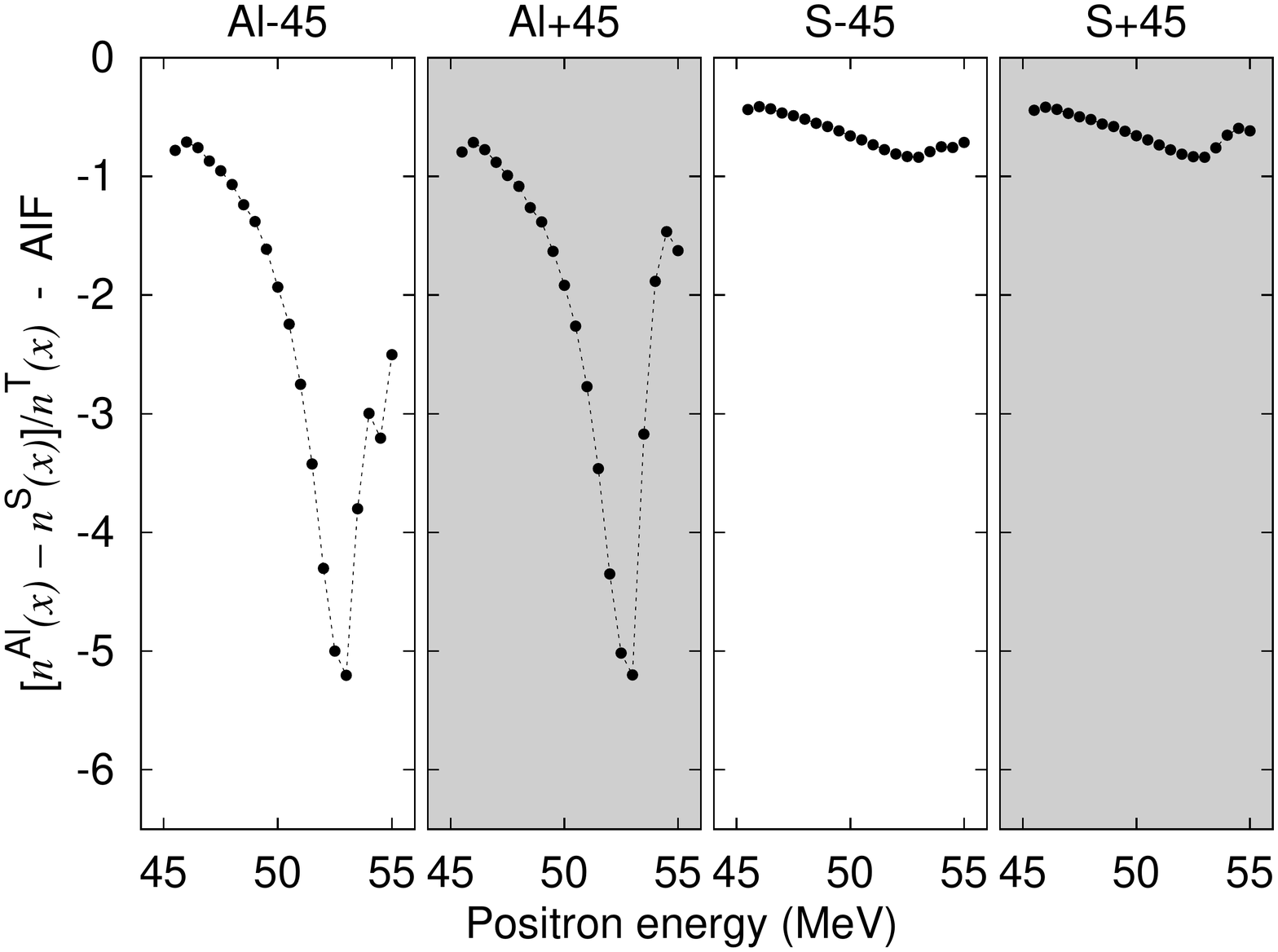}}
\centerline{\includegraphics[width=\linewidth]{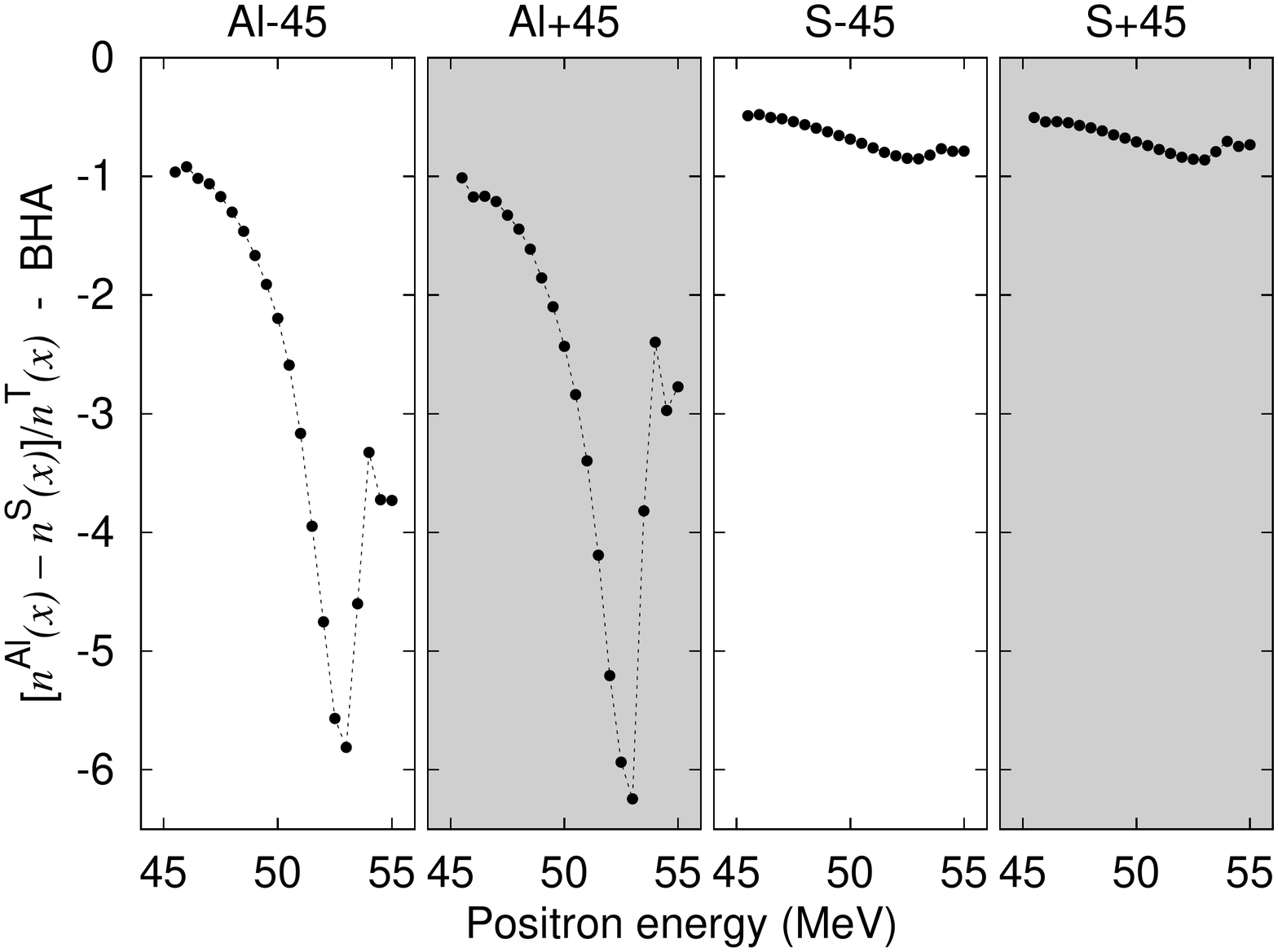}}
%vspace*{60mm}
\caption{Energy dependent part of the enhancement
  factor in Eq.(\ref{eq:enhfactT}) as deduced from the rates
  of AIF events (upper panel) and BHA events (lower panel) selected for
  the eight $\alpha$ configurations presented in
  Fig.~\ref{fig:superrat}. The enhancement factors for the two magnetization
  currents were averaged. The dotted lines between the points are to guide the eye.}
\label{fig:enh}
\end{figure}
 
Since the energy dependent part of the
enhancement factors for the two current polarities are very close, the
mean value, $k_\alpha$, has been taken as the enhancement factor used
in the final fit.
The results are shown in Fig.~\ref{fig:enh}.
The reduction of the enhancement factor observed at high
energy was anticipated by Monte-Carlo simulations performed during the
design of the spectrometer \cite{VanHove00} and is due to the finite
acceptance and momentum resolution.

We stress an important aspect of this experiment
using both polarized and depolarized muons, namely the fact 
that the enhancement factors are \emph{extracted} from the
transmitted positron rates measured with the
Al and S targets, as well as from a simple
determination of the corresponding muon polarization.
Since the very same device is used for the energy measurement and an identical
energy binning is applied to the extraction of the super-ratios and of the
enhancement factors, the data analysis does not require an accurate absolute
energy calibration of the spectrometer. Moreover, since the measurement of
the longitudinal polarization is differential, as a function of the positron
energy, the measurement does not require either an accurate determination
of the
absolute analyzing power of each scattering process selected by the polarimeter.

%%%%%%%%%%%%%%%%%%%%%%%%%%%%%%%%%%%%%%%%%%%%%%%%%%%%%%%%%%%%%%%%%%%%%%%%%%%%
\subsection{Fits}
\label{sec:fits}

According to Eq.~(\ref{eq:superrat3}), the super-ratios are
proportional to the positron polarization, $P_L$.
The dependence of $P_L$ as a function of $\Delta$ is given by
Eq.~(\ref{eq:longpol}) and is driven by the enhancement factors shown
in Fig.~\ref{fig:enh}.

As is visible in Fig.~\ref{fig:superrat}, the amplitudes of the
measured super-ratio are smaller than the maximal possible amplitudes
as given by the calculated functions ${\varphi(A^\pm_{i,\alpha}})$.
Two main sources have be considered to explain such differences:
1) A reduction could arise from a smaller magnetization in the
Vacoflux foil. A smaller electron polarization in the foil would 
reduce the effective analyzing power. If this were the only effect
responsible for the observed reduction,
the factor would then be the same for AIF and BHA processes.
However, this is not observed to be the case in the
super-ratios (Fig.~\ref{fig:superrat}). 
In any event, such a reduction of the super-ratios due to a non-saturation
of the foils has no dependence on the positron energy;
2) The contribution of background events remaining after the software cuts
and the misidentifications of events due to tracking
inefficiencies of the MWPC and detection inefficiencies in the
hodoscope tend to reduce the amplitude of the super-ratios.
Although, strictly speaking, the probability for a positron to produce
background events, such as a double bremsstrahlung, is naturally
energy dependent, such dependence varies smoothly over the energy window
considered in this experiment.
Neither of these two sources appears to be able to mimic 
a strong energy dependence like the one shown by the enhancement factors.
 
As a next step, in order to search for a possible energy dependence of the
longitudinal polarization, the super-ratios have then been fitted by the
function,
\begin{equation}
s_\beta(x) = a_\beta P_L(x)  \varphi(A_{\alpha}) + b_\beta\,,
\label{eq:superratfit1}
\end{equation}
where $a_\beta$ is a common attenuation factor for each pair of
configurations $\alpha$ associated with the $\pm 45^\circ$ orientations
of the foils, and $b_\beta$ is an offset for the same pair of
configurations.
The central assumption of this model is that the only
energy dependent behavior of the super-ratios is expected to arise
from the longitudinal polarization via the
enhancement factors. Each pair can otherwise have different
attenuation factors and offsets.

Replacing $P_L(x)$ by its expression after integration of the rates
over the spectrometer acceptance leads to
\begin{equation}
s_\beta(x) =
 a_\beta \xi' \left[ 1 + k_\alpha (P_\mu^T,x)~\Delta \right]  \varphi(A_{\alpha})
 + b_\beta\,.
\label{eq:superratfit2}
\end{equation}

The values of the muon polarization used in Eq.~(\ref{eq:enhfactT})
were $P_\mu^{\mathrm{Al}} = 0.937$ and $P_\mu^{\mathrm{S}} = 0.382$. 
Nine parameters were left free to fit all data: the four reduction
factors $a_\beta$, the four offsets $b_\beta$, and $\Delta$,
with the value of $\xi'$ fixed to $1$ \cite{PDG12}.

\begin{figure}[t]
\centerline{\includegraphics[width=\linewidth]{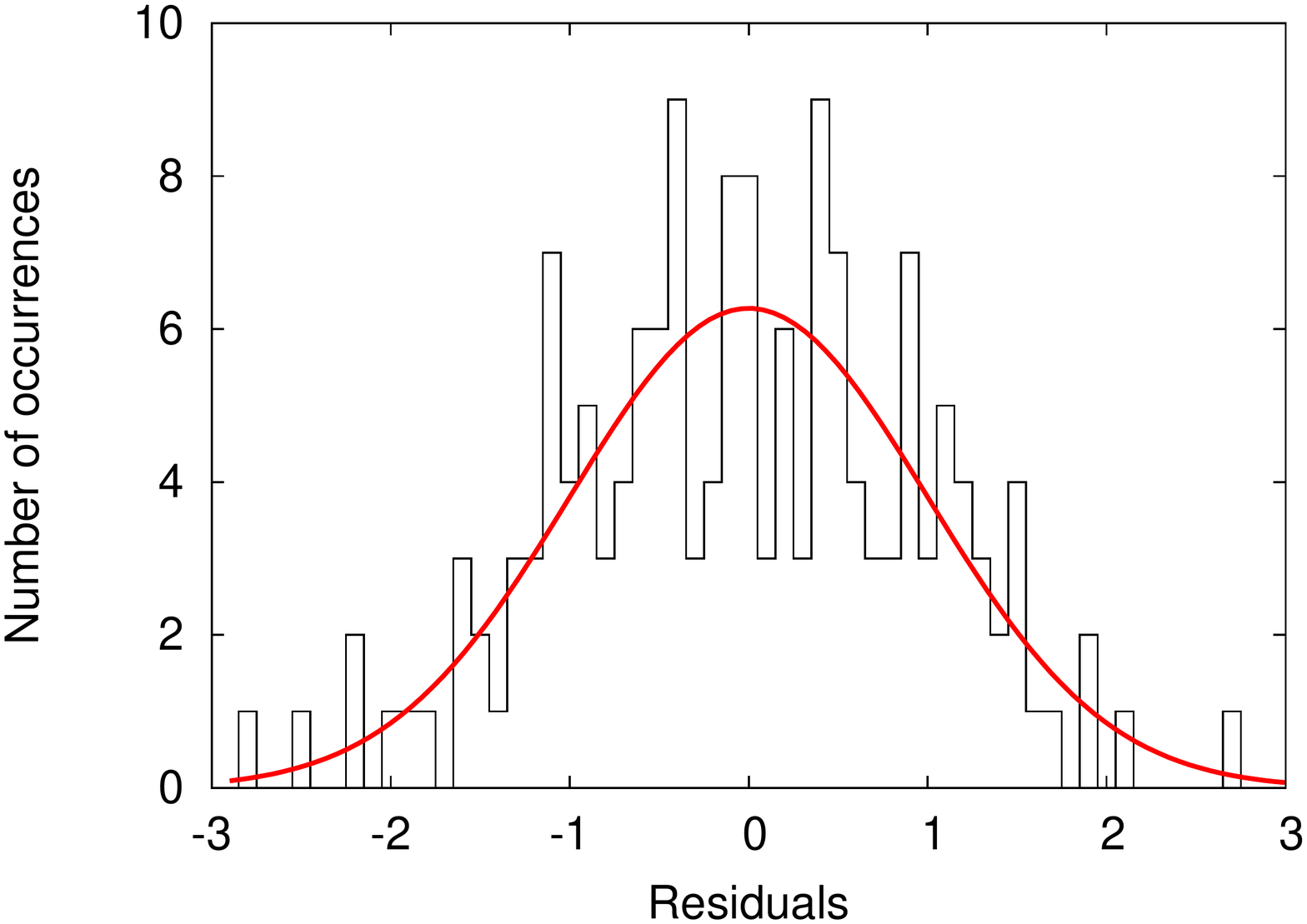}}
\vspace*{-3 mm}
\caption{(Color online) Distribution of residuals normalized to their statistical
  error as obtained from a fit of the super-ratios by a function with
  nine free parameters. The red curve is the fit of the residuals by a normal
  distribution.}
\label{fig:superratfit}
\end{figure}

The solid black
lines in Fig.~\ref{fig:superrat} show the results from the fit.
Figure~\ref{fig:superratfit} shows the residuals between the fit and
the data points, normalized to their statistical error.
Table~\ref{tab:delta} lists the values obtained for the fitted parameters
and their associated uncertainties.
The value of $\Delta$ obtained from the fit is $\Delta = -0.019(42)$ with
a reduced $\chi^2$ of $\chi^2/\nu = 1.17$.

\begin{figure}[t]
\centerline{\includegraphics[width= \linewidth]{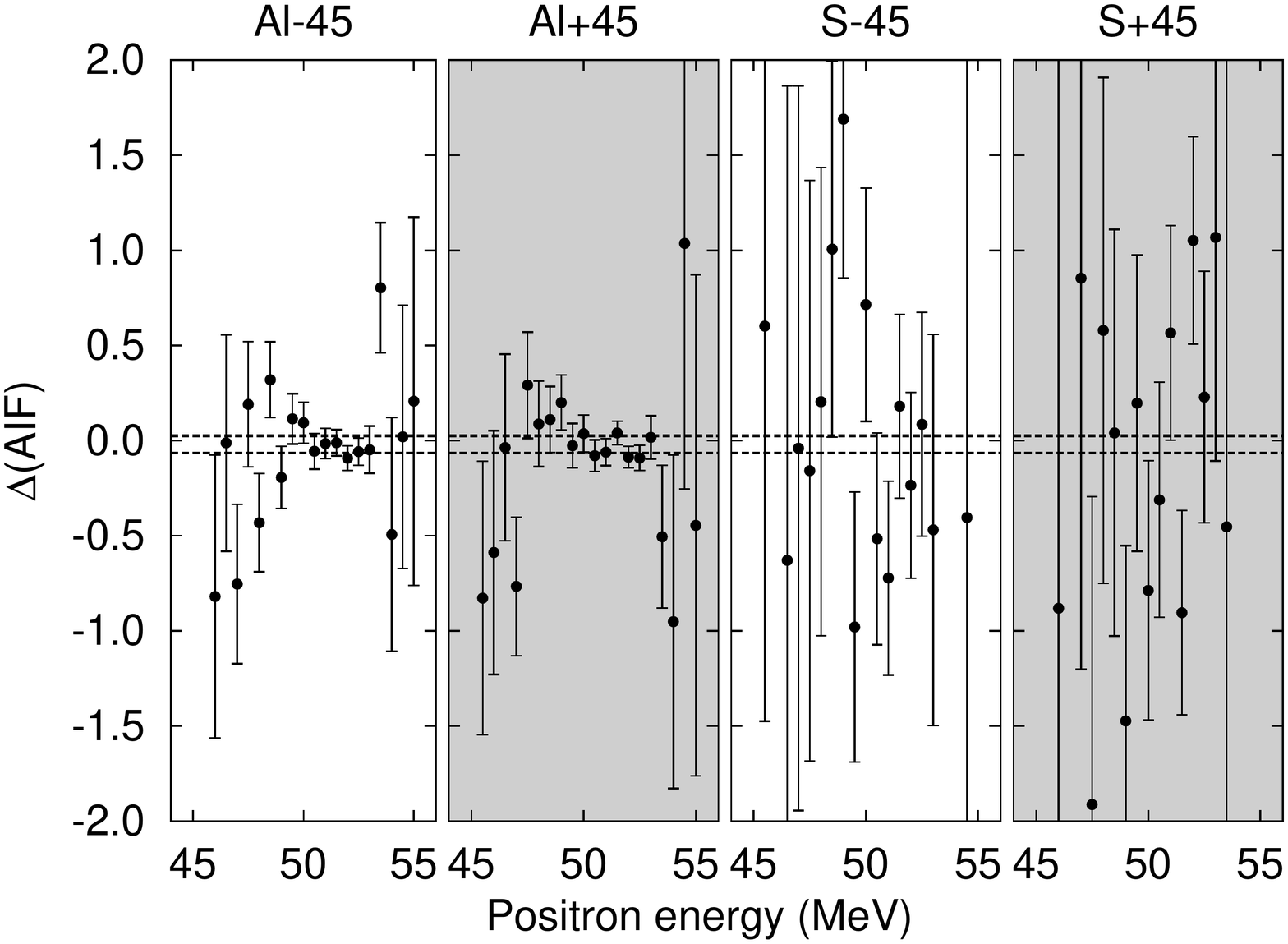}}
\centerline{\includegraphics[width=\linewidth]{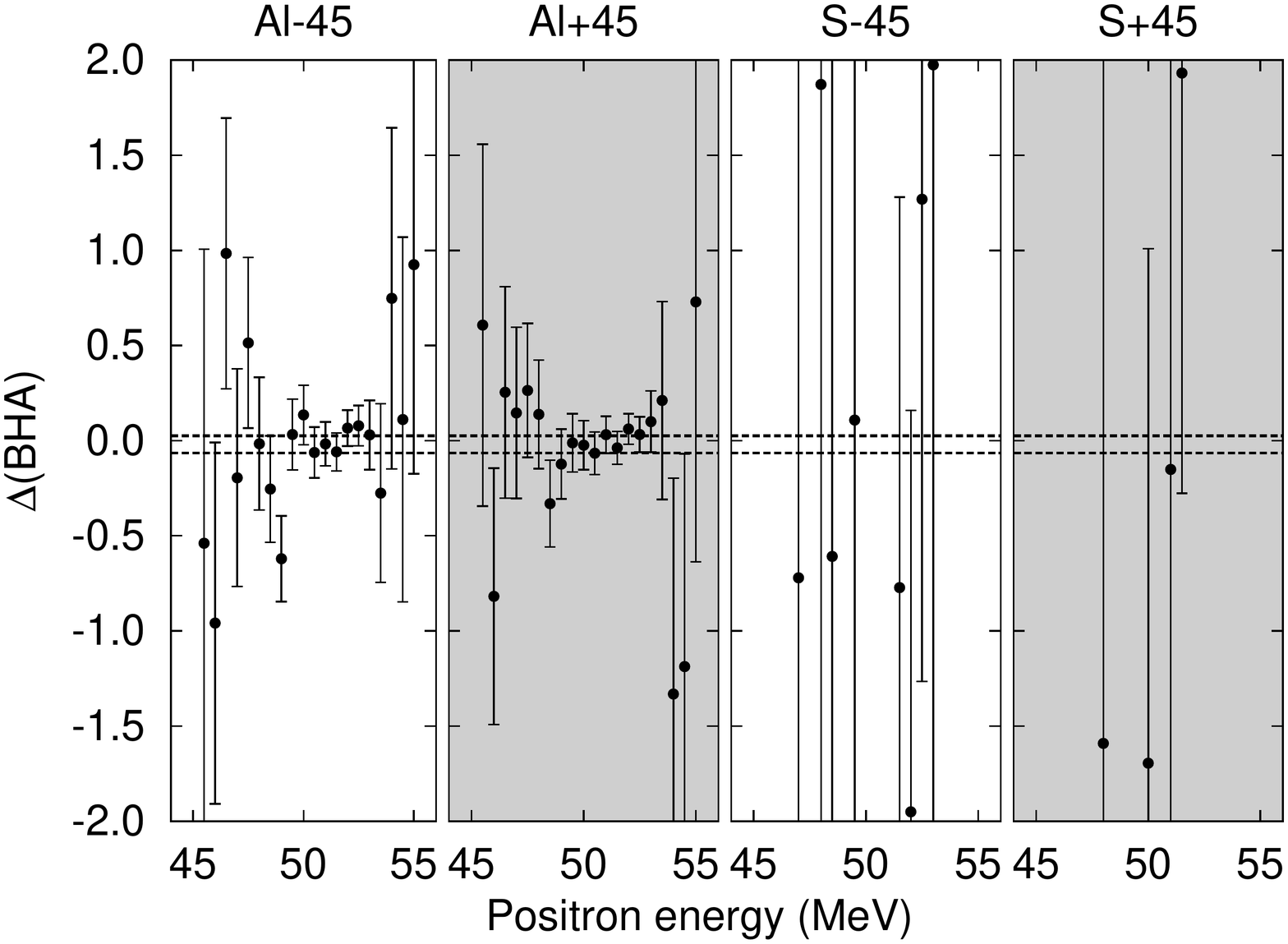}}
%vspace*{60mm}
\caption{The values of $\Delta$ with their errors obtained after
  inversion of Eq.~(\ref{eq:superratfit2}) for AIF events (upper panel)
  and BHA events (lower panel). The band delimited by the dotted lines
  indicate the $\pm1\sigma$ interval of the fitted value of $\Delta$.}
\label{fig:deltarev}
\end{figure}

Equation~(\ref{eq:superratfit2}) can be inverted and solved to express
each value of $\Delta$ and its error as deduced from the value of the
super-ratio.
The result is shown in Fig.~\ref{fig:deltarev}.
The large reduction of uncertainty resulting from the larger
enhancement factors
near the maximal energies for the measurements with the
polarization preserving Al targets is clearly visible.
The loss of sensitivity for the data taken with the S target is also
obvious.
The data associated with AIF events have a larger sensitivity and
dominate the precision on the final value of $\Delta$, but BHA
scattering data also contribute.

\newcommand{\phm}{\phantom{-}}
\begin{table}[ht]
\caption{Results from the fit of the super ratios $s_\beta(x)$,
  using values of the muon
  polarization $P^{\mathrm{Al}}_\mu = 0.937$ and $P^{\mathrm{S}}_\mu =
  0.382$.}
\begin{center}
\begin{tabular}{l@{\hspace{5mm}}l@{\hspace{5mm}}c@{\hspace{5mm}}c} \hline\hline
Process & Target  & $a_\beta$ & $b_\beta$ \\  \hline
AIF     & Al      & $ \phm 0.316 (46)$ & $ \phm 0.000 (2)$ \\
AIF     & S       & $ \phm 0.432 (29)$ & $ \phm 0.005 (2)$ \\
BHA     & Al      & $ \phm 0.239 (46)$ & $ \phm 0.004 (2)$ \\
BHA     & S       & $ \phm 0.118 (33)$ & $ \phm 0.005 (2)$ \\ \hline\hline
%
%\begin{tabular}{l@{\hspace{5mm}}r@{\hspace{5mm}}r@{\hspace{5mm}}l@{\hspace{5mm}}l} \hline\hline
%Parameter& Process & Target  & Value \\  \hline
%$\Delta$ &         &         & $     -0.019(42)$ \\
%$ a$     & AIF     & Al      & $ \phm 0.316 (46)$ \\
%$ a$     & AIF     & S       & $ \phm 0.432 (29)$ \\
%$ a$     & BHA     & Al      & $ \phm 0.239 (46)$ \\
%$ a$     & BHA     & S       & $ \phm 0.118 (33)$ \\
%$ b$     & AIF     & Al      & $ \phm 0.000 (2)$ \\
%$ b$     & AIF     & S       & $ \phm 0.005 (2)$ \\
%$ b$     & BHA     & Al      & $ \phm 0.004 (2)$ \\
%$ b$     & BHA     & S       & $ \phm 0.005 (2)$ \\  \hline\hline
%$\chi^2/\nu$   & $ \phm 1.17$  \\    \hline\hline
\end{tabular}
\end{center}
\label{tab:delta}
\end{table}

%%%%%%%%%%%%%%%%%%%%%%%%%%%%%%%%%%%%%%%%%%%%%%%%%%%%%%%%%%%%%%%%%%%%%%%%%%%%
\subsection{Residual muon polarization}
\label{sec:PS}

The preliminary measurements described in Sec.~\ref{sec:Pmu} indicated
that the residual polarization of muons in the S powder target could
be as low
as $P^{\mathrm{S}}_\mu=0.10(5)$, as obtained from the Hanle signals.
However, such a low
value of the residual polarization is not consistent with the values
obtained from the fits of the shape of the energy distributions as
measured with the spectrometer (Fig.~\ref{fig:pSpec}) during the main
experiment nor with the total positron yield ratio between
Al and  S targets normalized to the telescopes. 
In addition, the
values obtained for $P^{\mathrm{S}}_\mu$ from the fits of the energy distributions for
AIF, BHA, and MIC events are also not statistically compatible between
them although the $\chi^2$ distributions present rather flat minima.

We have adopted a conservative assumption by considering a
sufficiently broad interval for the residual polarization in the S target, 
$P^{\mathrm{S}}_\mu = 0.382(33)$, which is deduced
from the fits of the energy distributions for all configurations.
The impact of the values of the residual
polarization in the S target is then considered as a common (energy
independent) systematic effect. 
A similar procedure was applied to the positron energy distributions 
obtained with the Al target and resulted in a residual polarization
$P^{\mathrm{Al}}_\mu = 0.937(3)$.

Similar fits as the one described in Sec.~\ref{sec:fits} have been
performed with the extreme values $P^{\mathrm{S}}_\mu = 0.349$ and
$P^{\mathrm{S}}_\mu = 0.415$. 
The half difference between the central values of $\Delta$ obtained
from those fits plus the half
difference between the errors on $\Delta$
is then taken as an estimate of the systematic error associated with
the actual residual polarization in the S target.

The uncertainty on the muon polarization in the Al target has a
negligible effect on the final result.

%%%%%%%%%%%%%%%%%%%%%%%%%%%%%%%%%%%%%%%%%%%%%%%%%%%%%%%%%%%%%%%%%%%%%%%%%%%%
\subsection{Result}
\label{sec:result}

%The value of $\Delta$ for $P^{\mathrm{S}}_\mu = 0.382$ resulting from the
%fit is given on Table~\ref{tab:delta}.
Increasing the statistical error 
given in Sec. \ref{sec:fits} above by $\sqrt{\chi^2/\nu}= 1.08$ to account
for the quality of the fit and including the systematic error due to
the value of $P^{\mathrm{S}}_\mu$ as described above,
results in
\begin{equation}
\Delta = (-19 \pm 45_{\rm stat} \pm 3_{\rm syst}) \times 10^{-3}\,.
\label{eq:Delta-result}
\end{equation}
The uncertainty is dominated by statistics and the size of the systematic
error shows the sensitivity of the result to the determination of the
muon polarization.

\begin{figure}[!ht]
\centerline{\includegraphics[width=0.9\linewidth]{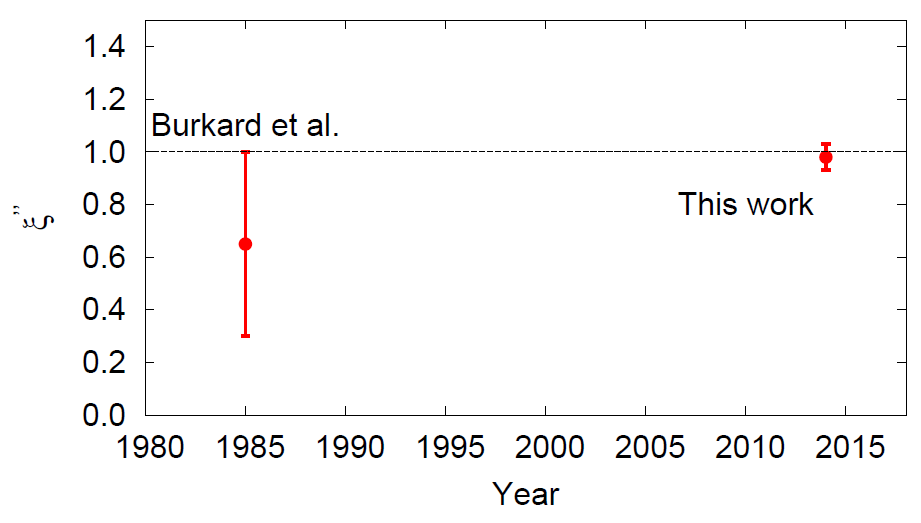}}
\caption{(color online) Measurements of the Michel Parameter $\xi''$ as a function of
the year of publication.}
\label{fig:xippHistory}
\end{figure}

From the definition of $\Delta$, Eq.~(\ref{eq:delta}), and
setting $\xi = \xi' = 1$ we get the value for $\xi''$:
\begin{equation}
\xi'' =0.981\pm 0.045_{\rm stat}\pm 0.003_{\rm syst}\,.
\label{eq:Xipp}
\end{equation}
This value is consistent with the SM expectation
$\xi^{\prime\prime}_{SM}=1$ and represents an
order of magnitude improvement (Fig.~\ref{fig:xippHistory}) on the relative
error over the current value
of $\xi''$ obtained under the same assumptions $\xi = \xi' = 1$
\cite{Burkard85a,Burkard85b}.

%%%%%%%%%%%%%%%%%%%%%%%%%%%%%%%%%%%%%%%%%%%%%%%%%%%%%%%%%%%%%%%%%%%%%%%%%%%%
\section{Implication on exotic couplings}

The combination of Michel parameters contained in $\Delta$
can be expressed in terms of the effective couplings which appear
in the interaction term, Eq.~(\ref{eq:mumatrix}).
The exact expression reads \cite{Fetscher95}
\begin{equation}
\Delta =
\frac{(a + 4b + 6c)(3a + 4b - 14c)}{(3a' + 4b' - 14c')(a' + 4b' + 6c')} - 1
\label{eq:Delta-abc}
\end{equation}
where $a$, $b$, $c$, $a'$, $b'$, and $c'$ are bilinear functions of the
couplings $g^{\gamma}_{\epsilon\mu}$ and are given in Refs.~\cite{Fetscher95,PDG12}.
Expanding to second order in the couplings which vanish in the SM, and
setting $g^V_{LL} = 1$, the expression in Eq.~(\ref{eq:Delta-abc}) becomes
\begin{eqnarray}
\Delta \approx &
8\left|g^V_{RL}\right|^2 + 4\left|g^V_{RR}\right|^2
+ \left|g^S_{RR}\right|^2 + \nonumber \\ 
&
+ 16\left|g^T_{RL}\right|^2
+ 8 {\rm Re}\left(g^S_{RL}g^{T*}_{RL}\right)\,.
\label{eq:Delta-gs}
\end{eqnarray}

Note that this quantity is sensitive to any exotic interaction which
would couple to the electron component of right-handed chirality.
%
% Wulf's comment cancels this fragment from the end of the above:
%  ... or the positron component of left-handed chirality. 
%
This includes the scalar, vector and tensor
interactions. The nature of the effective couplings to which this measurement is
sensitive is then different than other Michel parameters \cite{Bueno11,Hillairet12}
so that within the most general context of purely leptonic weak interactions
this experiment is definitely complementary to those measuring the spectrum shape
and the decay asymmetry.

A recent global analysis of muon decay data~\cite{Hillairet12} has provided new
limits on several of the couplings entering the expression of $\Delta$
in Eq.~(\ref{eq:Delta-gs}).
Considering the current 90\% C.L. limits $|g^V_{RR}| < 0.017$ and
$|g^S_{RR}| < 0.035$~\cite{Hillairet12}, we neglect here their contribution.
Further, we assume that time reversal invariance holds for all interactions
so that all couplings are taken to be real.

\begin{figure}[ht]
\centerline{\includegraphics*[width=\linewidth]{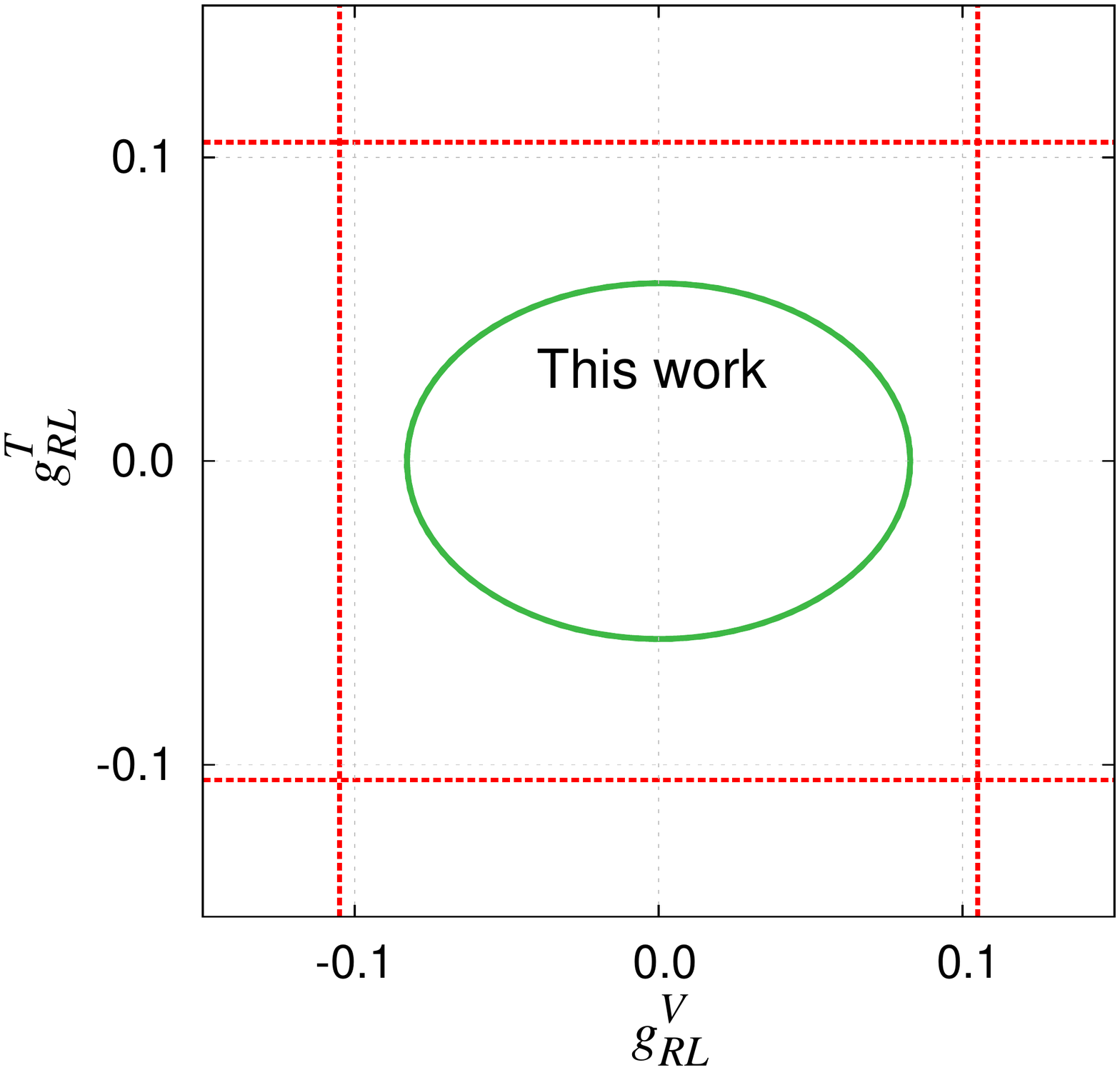}}
\vspace*{-3mm}
\caption{(Color online) Constraints at 90\% C.L. on exotic couplings
  $g^T_{RL}$ and $g^V_{RL}$ obtained from the present experiment
  (solid green curve), assuming $g^V_{RR} = g^S_{RR} = g^S_{RL} = 0$,
  and compared to the current limits (dashed red lines).}
\label{fig:excplot1}
\end{figure}

\begin{figure}[ht]
\centerline{\includegraphics[width=\linewidth]{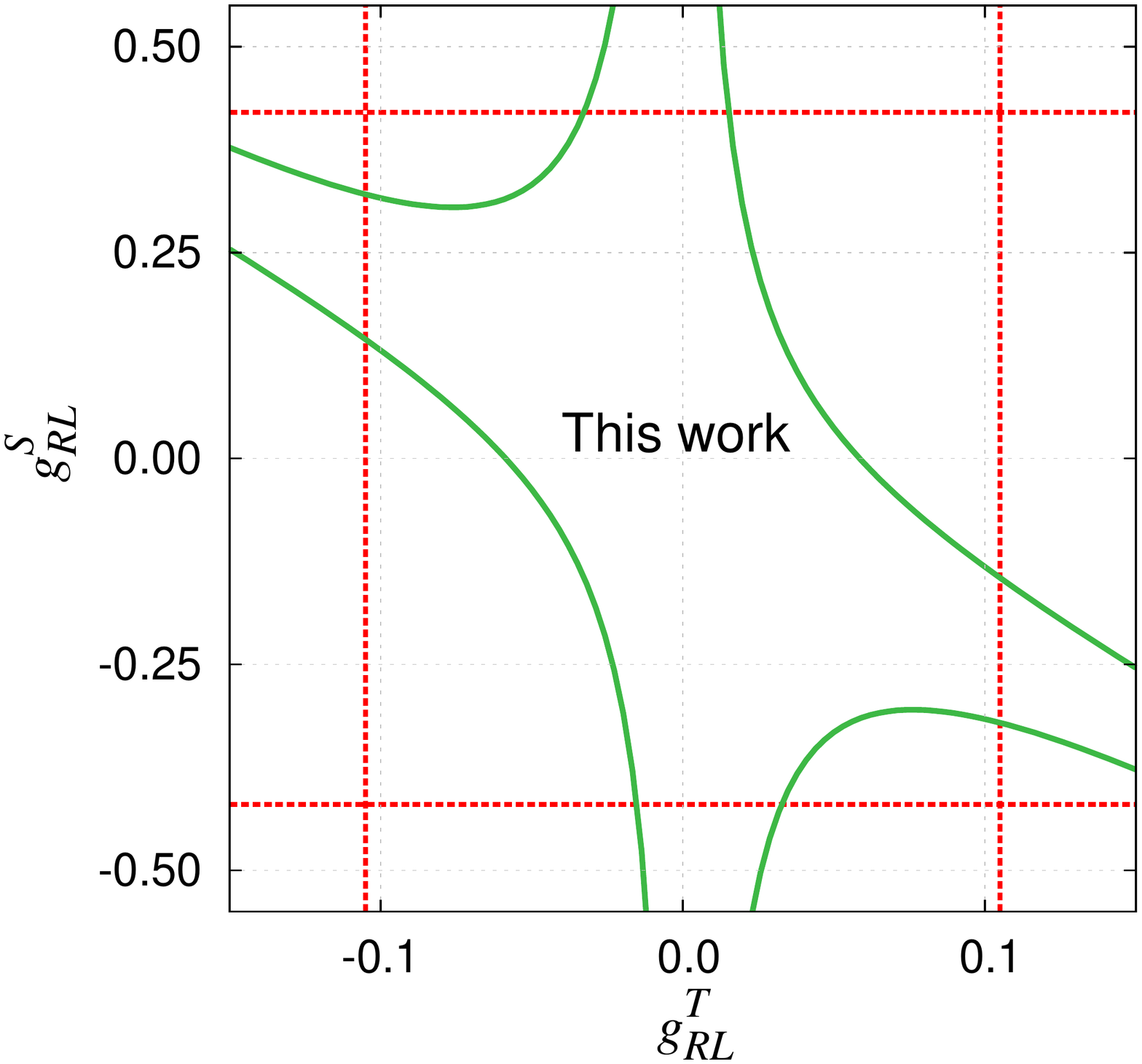}}
\vspace*{-3mm}
\caption{(Color online) Constraints at 90\% C.L. on exotic couplings
  $g^T_{RL}$ and $g^S_{RL}$ obtained from the present experiment
  (solid green curve), assuming $g^V_{RR} = g^S_{RR} = g^V_{RL} = 0$,
  and compared to the current limits (dashed red lines).}
\label{fig:excplot2}
\end{figure}

%  scenarii ??
%  http://english.stackexchange.com/questions/11775/what-is-the-plural-of-scenario
%  --> scenarios
%
To illustrate the level of sensitivity on exotic couplings obtained
from this measurement we select two scenarios, and provide two
dimensional exclusions plots, with either $g^S_{RL} = 0$ or $g^V_{RL}
= 0$.
Figure~\ref{fig:excplot1} shows the 90\% C.L. limits obtained on
$g^V_{RL}$ and $g^T_{RL}$ from this experiment (solid green line) as
compared to the current limits~\cite{Hillairet12} (dashed red lines).
The region outside the ellipse is excluded by the present work and
shows a significant reduction of the previously allowed parameter
region.
Figure~\ref{fig:excplot2} shows the 90\% C.L. constraints obtained on
$g^T_{RL}$ and $g^S_{RL}$ from this experiment (solid green line) as
compared to the current limits~\cite{Hillairet12} (dashed red lines).
The region outside the hyperbolas are excluded by the present work.

The result in Eq.~(\ref{eq:Delta-result}) can also be interpreted within
current specific scenarios extending the SM.
A natural framework for the interpretation of parity-violating (i.e.
pseudo-scalar) quantities is provided by left-right symmetric models
\cite{Mohapatra75}.
Such models introduce charged right-handed bosons, $W^\pm_R$, that
restore left-right symmetry by coupling to right-handed fermion
doublets.
The observation of parity violation at low energies is then attributed
to the large mass, $m_R$, of the right-handed bosons relative to the
standard left-handed one.

Several muon-decay parameters have been expressed in terms of
parameters of general left-right symmetric models \cite{Herczeg86}.
For the combination of Michel parameters which enter $\Delta$ in
Eq.~(\ref{eq:delta}) we have
\begin{equation}
\Delta = 4 v_e v_\mu r^4 \left(
\frac{\delta_M + t^2}{1 + \delta_M t^2} \right)^2
\label{eq:Delta-GLRM}
\end{equation}
where $v_l = \sum_i^\prime \left| U_{li}^R \right|^2 / \sum_i^\prime
\left| U_{li}^L \right|^2$ with $U_{li}^L$ ($U_{li}^R$) denoting the
elements of the Pontecorvo-Maki-Nakagawa-Sakata matrix,
coupling the charged left-(right)-handed lepton of flavour $l
= e,\ \mu$ to the mass eigenstate neutrino $\nu_i$; $r=g_R/g_L$ is the
ratio between the gauge couplings of the right-handed and left-handed
bosons; $\delta_M = (m_1/m_2)^2$, with $m_1$ ($m_2$) being the mass of
the light (heavy) boson; and $t = \tan \zeta$, with $\zeta$ the mixing
angle between the charged bosons $W_1$ and $W_2$.
The prime on the summation symbols indicates the inclusion of
neutrinos whose masses are sufficiently small that they couple to the
decay process.

To illustrate the sensitivity level to the heavy boson mass $m_2$, we
consider here the simple scenario of manifest left-right symmetry,
which implies $v_l = r = 1$.
Furthermore, including the tight constraint on the mixing angle,
$\zeta$, obtained from the unitarity condition of the
Cabibbo-Kobayashi-Maskawa quark mixing matrix \cite{Hardy09},
Eq.~(\ref{eq:Delta-GLRM}) reduces to
\begin{equation}
\Delta = 4\left( \frac{m_1}{m_2} \right)^4
\label{eq:Delta-MLRS}
\end{equation}
with $m_{1(2)} = m_{L(R)}$.
Under the assumptions above we find
\begin{equation}
m_R > 235 {~~\rm GeV}/c^2
\label{eq:mR}
\end{equation}
at 90\% C.L\@.
Such a mass scale is already excluded by other experiments in muon
decay \cite{Bueno11} as well as several direct and indirect searches
\cite{PDG12}.
This is however not surprising since, after all, the relative
precision on $\xi''$ obtained from this experiment is a moderate 5\%.

Independent of the above, and within the general phenomenological
description of the muon decay amplitude of Eq.~(\ref{eq:mumatrix}),
this experiment provides new model independent constraints on three of
the exotic
couplings which are neither accessible by recent high precision
measurements of other muon decay parameters nor by experiments at high
energy colliders.
A new global analysis of muon decay experiments, including the present
result and without making assumptions on specific couplings, would be
valuable in order to quantify the impact of all available data on the
couplings describing the leptonic weak interaction.

%%%%%%%%%%%%%%%%%%%%%%%%%%%%%%%%%%%%%%%%%%%%%%%%%%%%%%%%%%%%%%%%%%%%%%%%%%%%
\section{Summary}

We have provided a detailed description of the experimental setup and
of the analysis of a differential measurement of the longitudinal
polarization of positrons emitted from the decay of polarized and
depolarized muons.
The longitudinal polarization was measured as a function of the
positron energy near the maximum of the energy spectrum.
This property is sensitive to the Michel parameter $\xi''$ which has
previously been measured only once \cite{Burkard85a,Burkard85b}.

The development work and the preparation for this experiment were
carried out in 1995--2000 and the data presented here was accumulated
during a single six week run which took place in 2001.
From an early stage in the data analysis~\cite{Morelle02} it was
observed that the measured asymmetries for the two types of processes
were smaller than the maximum possible amplitudes expected in the case
the detected events were identified with full efficiency as pure
annihilation in flight or as Bhabha scattering.
Such smaller asymmetries, consistent with results from preliminary
Monte-Carlo simulations that included misidentified events, had been
observed in a preliminary test~\cite{VanHove00} performed in 1999.
The experiment was designed in such a way that Monte-Carlo simulations
are not absolutely necessary for the data analysis besides their
utility in the description of the spectrometer transmission function.
 
Despite the reduced sensitivity, the result obtained in this
measurement has improved the relative uncertainty of the Michel
parameter $\xi''$ by an order of magnitude, providing new constraints
of phenomenological couplings describing the leptonic weak interaction.

The uncertainty obtained from this measurement is dominated by the
statistical error which, in part, is also determined by the sensitivity.
It is conceivable that a future experiment could improve the
identification of the scattering events using tracking techniques
with detectors of lower mass. It is also possible to control the
residual muon polarization such as to
reduce the systematic error by at least a factor of 3, in order to
reach a precision level of $10^{-3}$ in a future improved measurement.

%%%%%%%%%%%%%%%%%%%%%%%%%%%%%%%%%%%%%%%%%%%%%%%%%%%%%%%%%%%%%%%%%%%%%%%%%%%%
\begin{acknowledgments}

We thank N.~Danneberg, M.~Hadri, C.~Hilbes, K.~K\"ohler, A.~Kozela,
Y.W.~Liu, R.~Medve, J.~Sromicki and F.~Foroughi  for their contributions
to the early phase of the experiment.
We are grateful to PSI for the excellent beam conditions, for the
loan of the PSC magnet (``Tracker'') as well as for the assistance of
the Hallendienst crew during the installation of the apparatus.
We express our gratitude to L.~Simons and B.~Leoni for the preparation
and use of the Cyclotron Trap magnet (``Filter'') and for their
support before and during the run.
We are greatly indebted to L.~Bonnet, P.~Demaret, and B.~de~Callata\"y
for
the development of SITAR (the SSD detector array) and to
A.~Ninane for the software assistance before and during the run.
This work was supported in part by the Belgian Institut
Interuniversitaire des Sciences Nucl\'eaires (IISN) and the Swiss
National Science Foundation (SNF).

\noindent
{\bf In memoriam}\\ 
We dedicate this work to our colleague and mentor, the late Professor
Jules P.~Deutsch, who passed away on February 5, 2011. He initiated
this work which still bears his style, insight, and influence. We
thank him deeply, and fondly keep the warmth of his memory.

\end{acknowledgments}
%%%%%%%%%%%%%%%%%%%%%%%%%%%%%%%%%%%%%%%%%%%%%%%%%%%%%%%%%%%%%%%%%%%%%%%%%%%%
%%%
\raggedright
\bibliography{composite}
%%%%%%%%%%%%%%%%%%%%%%%%%%%%%%%%%%%%%%%%%%%%%%%%%%%%%%%%%%%%%%%%%%%%%%%%%%%%
\end{document}